\documentclass[amsmath,twocolumn]{aastex702}

\shortauthors{B. Park et al.}
\graphicspath{{./}{figures/}}

\begin{document}

\title{Decomposition of Solar Wind Velocity Distribution Functions with Orthogonal Polynomials}

\author[0000-0002-7419-7999]{Byeongseon Park}
\affiliation{Departamento de Ciencias Espaciales, Instituto de Geof\'{i}sica, Universidad Nacional Aut\'{o}noma de M\'{e}xico}
\email[show]{bpark@igeofisica.unam.mx}

\author[0000-0002-0625-8892]{Primo\v{z} Kajdi\v{c}}
\affiliation{Departamento de Ciencias Espaciales, Instituto de Geof\'{i}sica, Universidad Nacional Aut\'{o}noma de M\'{e}xico}
\email{primoz@igeofisica.unam.mx}

\author[0000-0002-0497-1096]{Daniel Verscharen}
\affiliation{Mullard Space Science Laboratory, University College London, Dorking, RH5 6NT, UK}
\email{d.verscharen@ucl.ac.uk}

\author[0000-0001-7171-0673]{X\'{o}chitl Blanco-Cano}
\affiliation{Departamento de Ciencias Espaciales, Instituto de Geof\'{i}sica, Universidad Nacional Aut\'{o}noma de M\'{e}xico}
\email{xbc@igeofisica.unam.mx}

\author[0000-0003-4178-5206]{Jana {\v{S}}afr{\'{a}}nkov{\'{a}}}
\affiliation{Charles University, Faculty of Mathematics and Physics, Prague, Czech Republic}
\email{jana.safrankova@mff.cuni.cz}

\author[0000-0002-8160-3051]{Zden{\v{e}}k N{\v{e}}me{\v{c}}ek}
\affiliation{Charles University, Faculty of Mathematics and Physics, Prague, Czech Republic}
\email{zdenek.nemecek@mff.cuni.cz}

\author[0000-0001-8913-191X]{Alexander Pit{\v{n}}a}
\affiliation{Charles University, Faculty of Mathematics and Physics, Prague, Czech Republic}
\email{alex@aurora.troja.mff.cuni.cz}

\author[0000-0003-4247-4864]{Tereza {\v{D}}urovcov{\'{a}}}
\affiliation{Charles University, Faculty of Mathematics and Physics, Prague, Czech Republic}
\email{tereza.durovcova@mff.cuni.cz}

\author[0009-0001-0161-5192]{Sruti Satyasmita}
\affiliation{Charles University, Faculty of Mathematics and Physics, Prague, Czech Republic}
\email{sruti.satyasmita@matfyz.cuni.cz}

\author[0000-0002-2576-0992]{Jesse T. Coburn}
\affiliation{Laboratoire de Physique des Plasmas (LPP), CNRS, Observatoire de Paris, Sorbonne Université, Université Paris Saclay, École Polytechnique, Institut Polytechnique de Paris, 91120 Palaiseau, France}
\email{jesse.coburn@lpp.polytechnique.fr}

\begin{abstract}

We present a framework for decomposing solar-wind velocity distribution functions (VDFs) using orthogonal polynomial bases. We aim to establish a practical procedure for applying polynomial decompositions to \textit{in-situ} spacecraft VDFs and to clarify how the resulting spectra of expansion-coefficient power can be used for noise reduction, VDF reconstruction, and diagnostics of velocity-space structure. The method represents measured VDF structure with Hermite--Hermite and Hermite--Laguerre expansions, providing a nonparametric description of departures from Maxwellians, such as anisotropy, skewness, beams, and suprathermal tails. Expansion coefficients are estimated by Gaussian-weighted quadrature after interpolation of measured distributions onto polynomial nodes. We demonstrate several applications of polynomial decomposition to Solar Orbiter, Parker Solar Probe, and Magnetospheric Multiscale 1 measurements, including noise identification through high-order spectral flattening, noise-reduced VDF reconstruction, and characterization of VDF-structure variations under different plasma conditions, \textit{e.g.,} turbulent solar-wind streams and shocks. For instance, noise-reduced reconstructed VDFs can provide smoother estimates of distinct ion populations and VDF gradients. Examples from solar-wind streams and collisionless-shock crossings further show that the resulting spectra respond to changes in parallel and perpendicular VDF structure, illustrating their potential for comparing kinetic modifications under different plasma conditions. Overall, orthogonal-polynomial decomposition provides a bridge between measured particle distributions and kinetic plasma physics by converting complex VDF morphology into quantitative velocity-space spectra.

\end{abstract}

\section{Introduction}\label{sec_1}

Solar wind plasma consists of multiple particle species, \textit{viz.,} electrons, protons, alpha particles, and a much smaller abundance of heavier ions \citep{marsch2006,verscharen2019}. In collisional media that evolve toward thermal equilibrium, particle velocity or energy distributions often approach Gaussian (or Maxwellian) forms. However, solar wind plasma is weakly collisional with characteristic interparticle collision times exceeding the particle transit time to 1 au \citep{kasper2008,verscharen2019,mostafavi2024}. Consequently, velocity distributions generally exhibit non-Gaussian features, including skewness and tails, for which a simple Gaussian model is insufficient.

On the basis of solar wind velocity distributions, several populations can be identified. For example, ion populations comprise quasi-thermal proton and alpha-particle cores and suprathermal beams \citep{marsch2006,hellinger2013,durovcova2019,demarco2023}, whereas electron populations can be separated into quasi-thermal core, suprathermal halo, and strahl (German for ``beam'') components \citep{abraham2022,eyelade2025}. The innermost regions of proton, electron, and $\alpha$-particle core populations are often approximately described by Maxwellian or bi-Maxwellian distributions, although temperature anisotropy and other departures from thermal equilibrium are commonly observed \citep{stverak2009}. In contrast, because particles preferentially stream along magnetic-field lines, velocity distributions in space plasmas often exhibit skewness along the magnetic-field direction \citep{marsch1982,hellinger2011,louarn2024}. This effect produces distribution-function asymmetries that cannot be fully described by a Gaussian model.

Numerous analytical models have been developed for these suprathermal populations. \citet{louarn2024} introduce skewness and kurtosis into a Gaussian model for ion velocity distributions using a normal inverse Gaussian (NIG) distribution. In this framework, stochastic variations in particle drift and temperature generate asymmetric and heavy-tailed deviations from a Gaussian distribution, allowing the model to represent both skewness and kurtosis. The resulting skewed distribution with a high-energy tail associated with magnetic-field-aligned (MFA) dynamics can therefore be interpreted in terms of heating and acceleration processes.

Other parametric approaches have been developed to describe the suprathermal tails of electron velocity distribution functions (VDFs). The $\kappa$ distribution function has been widely adopted to model suprathermal electron halos \citep{livadiotis2018,pierrard2022}. This model contains pronounced tails than either the NIG or Gaussian distribution, controlled by the $\kappa$ index. Given that the electron core is approximately Maxwellian, a combined model consisting of a Gaussian core and a $\kappa$-distributed halo is also considered \citep{lazar2017}. However, like the Gaussian distribution, the $\kappa$ distribution is symmetric and therefore cannot adequately represent the asymmetry, or skewness, of observed distributions. In this context, \citet{zenteno2021} introduce skewness into the $\kappa$ distribution function with a second-order Taylor expansion.

The aforementioned parametric models describe particle behavior in a physically interpretable manner, \textit{e.g.,} population drifts, relative drifts between different populations, temperatures, heating, and acceleration \citep{araneda2002,marsch2006,hellinger2013}. However, the assumptions underlying these models, such as thermal equilibrium, and their fitting to observations inevitably involve model- and analyst-dependent choices. Machine-learning techniques, such as a Gaussian mixture model (GMM) with an expectation--maximization algorithm, have recently been applied to several topics in space physics with the aim of reducing such analyst-dependent factors \citep{demarco2023,edens2024,sano2025,ran2026}. However, when overlapping Gaussian distributions are not sufficiently separable, such as closely adjacent proton core and beam populations, the GMM may fail to identify them as distinct physical components \citep{sano2025}. Furthermore, when distributions exhibit a ``flat-top'' shape \citep{wilson2020}, the model may represent the structure using multiple superposed Gaussian functions \citep{dupuis2020}.

Direct decomposition of solar wind particle VDFs using a set of orthogonal polynomials, \textit{e.g.,} Hermite and Laguerre polynomials, provides a promising analogue of Fourier analysis for VDFs. Wave-like signals of the magnetic and electric fields can be decomposed by Fourier transforms into an effectively infinite set of orthogonal basis functions in Hilbert space with the corresponding coefficients yielding spectra. Spectral analyses of magnetic fields in space plasmas have been extensively used to study solar wind turbulence and dissipation \citep{alexandrova2008,Bruno2013,goldstein2015,smith2024}. In a similar manner, VDF decomposition computes the coefficients of polynomial basis functions and provides a spectrum that can be used to investigate the possible transfer of free energy toward progressively finer velocity-space structure \citep{servidio2017,pezzi2018,cerri2018,larosa2025}. This decomposition approach does not require fitting a prescribed model to VDFs; instead, it provides a nonparametric representation of VDF structure. It is therefore a promising technique for broader application in kinetic plasma studies.

The present study is complementary to previous applications of polynomial representations of VDFs. For instance, \citet{servidio2017} demonstrate that Hermite spectra can be used to diagnose velocity-space cascades, \citet{coburn2024} apply Hermite--Laguerre filtering to obtain smooth electron VDF gradients for wave--particle interaction analysis, and \citet{larosa2025} use Hermite spectra to investigate turbulent cascade in the velocity-space near the Sun. We focus on the practical construction and interpretation of the decomposition itself. In particular, we (a) compare Hermite--Hermite and Hermite--Laguerre representations, (b) discuss the effects of quadrature order, choice of polynomial basis, and instrumental resolution, and (c) provide a methodological guide for applying polynomial decomposition to spacecraft VDFs. This framework is then used to illustrate several applications, such as noise reduction, shock-related VDF modifications, and future diagnostics of kinetic wave--particle interactions.

In this paper, Section \ref{sec_2} describes the spacecraft data sets and preprocessing procedures, followed by the theoretical formulation and numerical implementation of VDF decomposition using Hermite and Laguerre polynomials. Their basic properties, including orthogonality, are discussed. Section \ref{sec_3} presents applications of the decomposition method to VDF measurements from, \textit{e.g.,} Solar Orbiter and Magnetospheric Multiscale missions, including noise identification, noise-reduced VDF reconstruction, and quantitative characterization of VDF variations across heliospheric shocks and turbulent solar-wind streams. Section \ref{sec_4} discusses the physical interpretation and broader aspects of this mathematical approach with its caveats.

\section{Polynomial decomposition of VDF}\label{sec_2}

The decomposition of VDFs is based on orthogonal polynomials, such as Hermite or Laguerre polynomials, in a manner analogous to Fourier analysis. The associated quadrature rule also requires the determination of polynomial roots, \textit{e.g.,} $x_i \in \{ x|H_m(x)=0\}$ for an $m$th-order Hermite polynomial, where $x_i$ denotes the $i$th node. This section first describes the spacecraft data sets and preprocessing procedures used in the analysis. We introduce the orthogonal-polynomial framework for VDF decomposition, including the relevant properties of Hermite and Laguerre polynomials and the calculation of their expansion coefficients.

\subsection{Data Set and Preprocessing}\label{sec_21}

In this study, the following spacecraft, instruments, and data products are used.

\begin{enumerate}
\item 
Solar Orbiter (SolO), Solar Wind Analyzer (SWA; \citealt{owen2020}), and Magnetometer (MAG; \citealt{horbury2020}) \footnote[1]{https://soar.esac.esa.int/soar/}. The primary ion VDFs used in Sections \ref{sec_2} and \ref{sec_3} are obtained from the Proton and Alpha Particle Sensor (PAS) of SWA. PAS samples 11 azimuth and 9 elevation angles with 96 energy channels spanning approximately $200~\mathrm{eV}$--$20~\mathrm{keV}$, producing $11\times9\times96$ three-dimensional (3D) VDF tensors in the instrument frame. A complete 3D VDF is acquired in about 1 s, while data product used in this study contains VDF measurements at a 4 s cadence. The example PAS distribution used for decomposition processes throughout this study was obtained at 2022 March 8 14:45:22 UT. It is selected from the vicinity of an interplanetary shock reported by \citet{trotta2024} because the VDF contains pronounced nonthermal particle populations formed by reflected particles from the shock front. Electron VDFs used for the reconstruction test in Appendix \ref{sec_g} are obtained from the Level 3 pitch-angle distribution data measured by the Electron Analyser System (EAS) of SWA, which consists of two sensors, EAS1 and EAS2, each sampling 16 elevations $\times$ 32 azimuths $\times$ 64 energy channels. The magnetic-field data used to define the MFA coordinates are provided by SolO/MAG at burst-mode cadences of 64--128 vectors $\mathrm{s}^{-1}$.

\item 
Magnetospheric Multiscale 1 (MMS1), Fast Plasma Investigation (FPI; \citealt{pollock2016}), and FIELDS/Fluxgate Magnetometer (FGM; \citealt{torbert2016}) \footnote[2]{https://lasp.colorado.edu/mms/sdc/public/about/browse-wrapper/}. To illustrate the application of the decomposition method to the Earth's bow shock, fast-mode ion VDFs from MMS1 are employed and presented in Section \ref{sec_32}. The FPI ion VDFs are processed as $32\times16\times32$ tensors, corresponding to 32 azimuth angles, 16 elevation angles, and 32 energy channels spanning approximately 10 eV--30 keV, with a cadence of 4.5 s. The corresponding magnetic-field data are obtained from MMS1/FGM at survey-mode cadences of 8--16 samples $\mathrm{s}^{-1}$.

\item 
Parker Solar Probe (PSP), Solar Wind Electrons Alphas and Protons (SWEAP; \citealt{kasper2016}) \footnote[3]{https://sweap.cfa.harvard.edu/pub/data/sci/sweap/}, and FIELDS/MAG (FIELDS; \citealt{bale2016}) \footnote[4]{https://research.ssl.berkeley.edu/data/psp/data/sci/fields/}. Appendix \ref{sec_f} compares polynomial spectra estimated from SolO and PSP observations. For PSP, we use ion VDFs from the Solar Probe Analyzers (SPAN) of SWEAP. The SPAN-I data product considered here samples 8 elevations $\times$ 8 azimuths $\times$ 32 energy channels, with a nominal cadence of approximately 7 s. The magnetic-field data are provided by the fluxgate magnetometers of PSP/FIELDS, with cadences ranging from 2.29 to 292.9 samples $\mathrm{s}^{-1}$, depending on the data product.
\end{enumerate}

All VDFs are transformed from the instrument frame into an MFA coordinate system. Its basis is defined using the local mean magnetic-field ($\mathbf B$) direction, $\hat{\mathbf e}_{\parallel}=\hat{\mathbf b}=\mathbf B/|\mathbf B|$, the first perpendicular direction, $\hat{\mathbf e}_{\perp1}=(\mathbf V_{\rm SW}\times\hat{\mathbf b})/|\mathbf V_{\rm SW}\times\hat{\mathbf b}|$ (where $\mathbf V_{\rm SW}$ denotes the solar-wind bulk velocity derived from moment analysis), and the second perpendicular direction, $\hat{\mathbf e}_{\perp2}=\hat{\mathbf e}_{\parallel}\times\hat{\mathbf e}_{\perp1}$. The particle velocities are then projected onto these three directions to obtain $v_{\parallel}$, $v_{\perp1}$, and $v_{\perp2}$.

To enable comparison among VDFs measured at different times, by different instruments, and under different plasma conditions, the velocity coordinates are shifted so that the proton-core bulk velocity defines the origin of velocity space, while the thermal speeds set its characteristic scales. To determine these reference quantities, we first fit a global 3D Maxwellian function to measured VDFs defined as

\begin{equation}
    \begin{split}
    f_{c,M}(v_{\parallel},v_{\perp1},v_{\perp2}) =
    \frac{n}{\pi^{3/2} w_{c,\parallel}w_{c,\perp}^2}
    \exp \biggl[
    -\frac{(v_{\parallel} - v_{c,\parallel})^2}{w_{c,\parallel}^2}
    \\
    - \frac{(v_{\perp 1} - v_{c,\perp 1})^2}{w_{c,\perp}^2}
    - \frac{(v_{\perp 2} - v_{c,\perp 2})^2}{w_{c,\perp}^2}
    \biggr],
    \end{split}
    \label{3dmax}
\end{equation}

\noindent where $n$ is the number density, $v_{c,\parallel}$ is the parallel, while $v_{c,\perp1}$ and $v_{c,\perp2}$ are the two perpendicular drift velocities of the proton core, and $w_{c,\parallel}$ and $w_{c,\perp}$ are the parallel and perpendicular proton-core thermal speeds, respectively. The perpendicular drift velocities are retained, although their values are much smaller than $v_{c,\parallel}$ in the MFA frame, so that the same fitting framework can be consistently extended to drifting secondary populations, such as the proton beam and $\alpha$ particles, in Section \ref{sec_31} \footnote[5]{Electrostatic analyzers such as SWA-PAS and SPAN-I measure ions in energy-per-charge and look-direction; conversion to MFA velocity coordinates can therefore introduce small perpendicular offsets that do not necessarily represent physical cross-field drifts \citep{owen2020,kasper2016}.}. The normalized velocities are then defined as

\begin{equation}
    \xi_{\parallel} =
    \frac{v_{\parallel}-v_{c,\parallel}}{w_{c,\parallel}},
    \qquad
    \xi_{\perp1,2} =
    \frac{v_{\perp1,2}-v_{c,\perp1,2}}{w_{c,\perp}}.
    \label{vnml}
\end{equation}

\noindent With this normalization, the fitted proton core is centered at the origin of normalized velocity space. 

For the two-dimensional (2D) gyrotropic representation, we use

\begin{equation}
    \xi_{\perp} = \sqrt{\xi_{\perp1}^{2}+\xi_{\perp2}^{2}}.
    \label{gytrpy}
\end{equation}

\noindent This reduction assumes that the dominant large-scale VDF structure depends primarily on $v_{\parallel}$ and $v_{\perp}$, while variations in gyrophase are averaged over. The reduced 2D Maxwellian or bi-Maxwellian form derived from equation \ref{3dmax} is also used to estimate beam and $\alpha$-particle populations in Section \ref{sec_31}.

\subsection{Orthogonal Basis Functions}\label{sec_22}

The basic principle of the decomposition is to represent a measured VDF as a weighted sum of basis functions, where each coefficient quantifies the contribution of a particular velocity-space mode. Orthogonality is essential because it allows each coefficient to be determined independently by projection --- thereby suppressing redundancy between modes --- and permits the coefficient power to be interpreted as a spectral measure of velocity-space structure. 

Although other expansions are possible, such as Fourier or Taylor representations, they are less naturally suited to spacecraft VDFs; Fourier bases impose artificial periodicity on nonperiodic velocity-space distributions, while Taylor expansions are generally nonorthogonal. Hermite and Laguerre functions are preferred here as they form orthogonal bases containing Gaussian kernels, $e^{-x^2}$, so that deviations from the Gaussian distribution are represented by a set of these polynomials, on domains appropriate for VDF analysis; Hermite functions describe the coordinates of $(-\infty,\infty)$, whereas Laguerre functions describe positive radial variables such as $\mu=\xi_{\perp}^{2}$ on $[0,\infty)$.

The orthogonality of the Hermite basis is expressed as

\begin{equation}
    \int^{\infty}_{-\infty} \eta_m(x) \eta_n(x) dx = \delta_{mn},
    \label{eq2}
\end{equation}

\noindent where the Hermite basis is

\begin{equation}
    \eta_m(x)=\frac{1}{\sqrt{2^{m}m!\sqrt{\pi}}}e^{-x^2/2} H_m(x),
    \label{eq3}
\end{equation}

\noindent and the $m$-order Hermite polynomial $H_m(x)$ is defined as

\begin{equation}
    H_m(x)=(-1)^{m} e^{x^2} \frac{d^m}{dx^m} e^{-x^2}.
    \label{eq4}
\end{equation}

\ The orthogonality of the Laguerre basis is written as

\begin{equation}
    \int^{\infty}_{0} \lambda_m^k(x) \lambda_n^k(x) dx = \delta_{mn},
    \label{eq5}
\end{equation}

\noindent where the Laguerre basis is

\begin{equation}
    \lambda_m^k(x)=\sqrt{\frac{m!}{(m+k)!}} e^{-x/2} x^{k/2} L_m^k(x),
    \label{eq6}
\end{equation}

\noindent and the associated Laguerre polynomial of degree $m$ and parameter $k$, $L_m^k(x)$ is defined as

\begin{equation}
    L_m^k(x)=\frac{e^x x^{-k}}{m!} \frac{d^m}{dx^m} (e^{-x} x^{m+k}).
    \label{eq7}
\end{equation}

\noindent When $x^2$ is used as the argument of the Laguerre basis, the associated Laguerre function contains, $L_m^k(x^2)$, the Gaussian kernel. Throughout this study, we set $k=0$ because the reduced gyrotropic VDF is represented only in terms of the radial perpendicular velocity $\mu=\xi_{\perp}^{2}$, without additional angular dependence or extra radial weighting. Nonzero values of $k$ would define a different associated-Laguerre basis with an additional $\mu^k$ weighting, which is not required for the present decomposition.

\subsection{Polynomial Representations of VDFs}\label{sec_23}

A VDF $f(\xi_{\parallel},\xi_{\perp})$ can be represented as a sum of polynomial basis functions,

\begin{equation}
    f^{HH}(\xi_{\parallel},\xi_{\perp})=\sum_{m=0}^{M} \sum_{n=0}^{N} c_{mn}^{HH} \eta_{m}(\xi_\parallel) \eta_{n}(\xi_\perp),
    \label{eq8}
\end{equation}

\noindent for Hermite polynomials in both the parallel and perpendicular directions. Hereafter, this decomposition is referred to as the Hermite--Hermite, or HH, representation. $f^{HH}$ indicates a VDF represented by HH decomposition, the quantity $c_{mn}^{HH}$ denotes the HH coefficient for polynomial orders $m$ and $n$, and $M$ and $N$ are the maximum polynomial orders used in the analysis.

The Laguerre polynomial can instead be used for the perpendicular component, yielding the Hermite--Laguerre, or HL, representation,

\begin{equation}
    f^{HL}(\xi_{\parallel},\sqrt{\mu})=\sum_{m=0}^{M} \sum_{n=0}^{N} c_{mn}^{HL} \eta_{m}(\xi_\parallel) \lambda_{n}^0(\mu),
    \label{eq9}
\end{equation}

\noindent where $f^{HL}$ is a VDF of HL representation, $c_{mn}^{HL}$ is the coefficient of the $m$th- and $n$th-order HL basis functions. The polynomial coefficients are estimated as

\begin{equation}
    c_{mn}^{HH} = \int_{-\infty}^{\infty} \int_{-\infty}^{\infty} f^{HH}(\xi_{\parallel},\xi_{\perp}) \eta_{m}(\xi_\parallel) \eta_{n}(\xi_\perp) d\xi_\parallel d\xi_\perp,
    \label{eq10}
\end{equation}

\noindent for the HH polynomials, and as

\begin{equation}
    c_{mn}^{HL} = \int_{-\infty}^{\infty} \int_{0}^{\infty} f^{HL}(\xi_{\parallel},\sqrt{\mu}) \eta_{m}(\xi_\parallel) \lambda_{n}^0(\mu) d\xi_\parallel d\mu,
    \label{eq11}
\end{equation}

\noindent for the HL representation, for each pair of orders $m$ and $n$. 

The 2D coefficient estimates in equations \ref{eq10} and \ref{eq11} are based on the gyrotropy assumption. \citet{larosa2025} assume $f(\xi_{\perp})=f(-\xi_{\perp})$; therefore, HH polynomials, which span $(-\infty,\infty)$ in both the parallel and perpendicular directions, are used. When HL representation is considered, the Laguerre polynomial spanning $[0,\infty)$ replaces the perpendicular Hermite basis used in the HH expansion. This choice corresponds to cylindrical coordinates with $\xi_{\perp}$ acting as a radial coordinate, in contrast to the Cartesian-like coordinates used for the HH representation, because $d\mu=2\xi_{\perp}d\xi_{\perp}$.

For 3D VDF measurements, equations \ref{eq8} and \ref{eq10} can be generalized directly using three Hermite polynomial bases, hereafter denoted H3, without imposing the gyrotropy:

\begin{equation}
    f^{H3}(\xi_{\parallel},\xi_{\perp1},\xi_{\perp2})=\sum_{m=0}^{M} \sum_{n=0}^{N} \sum_{l=0}^{L} c_{mnl}^{H3} \eta_{m}(\xi_\parallel) \eta_{n}(\xi_{\perp1}) \eta_{l}(\xi_{\perp2}),
    \label{eqh31}
\end{equation}

\noindent and

\begin{equation}
    \begin{split}
    c_{mnl}^{H3} = \int_{-\infty}^{\infty} \int_{-\infty}^{\infty} \int_{-\infty}^{\infty} f^{H3}(\xi_{\parallel},\xi_{\perp1},\xi_{\perp2}) \times \\
    \eta_{m}(\xi_\parallel) \eta_{n}(\xi_{\perp1}) \eta_{l}(\xi_{\perp2}) d\xi_\parallel d\xi_{\perp1} d\xi_{\perp2},
    \end{split}
    \label{eqh32}
\end{equation}

\noindent where $f^{H3}$ is a VDF represented by the H3 approach and $c_{mnl}^{H3}$ is the coefficient of the $m$th-, $n$th-, and $l$th-order Hermite basis functions. 

Although a full 3D Hermite decomposition can be applied to the VDF, the present study focuses mainly on 2D reduced representations because they provide a practical framework for gyrotropic solar-wind VDFs. A 3D decomposition retains nongyrotropic structure and can reduce some interpolation issues associated with projecting measurements onto the $(\xi_{\parallel},\xi_{\perp})$ plane. However, it is computationally more expensive, less straightforward to visualize, and still does not exactly match the native energy--angle coordinates of top-hat plasma instruments. A comparison between 2D and 3D Hermite decompositions is discussed in Appendix \ref{sec_h}.

The HH representation is useful as a simple diagnostic basis as the same Hermite functions are used in the parallel and perpendicular directions, allowing direct comparison of parallel and perpendicular coefficients of the spectrum. The HL representation is generally more appropriate for reduced gyrotropic VDFs as the perpendicular coordinate is represented by the positive radial variable, $\mu$, and expanded with Laguerre functions, avoiding the artificial mirroring of the perpendicular direction.

To compute the expansion coefficients, the continuous integrals are evaluated using Gaussian-weighted quadrature (Appendix \ref{sec_b}). This requires three main ingredients: the corresponding quadrature nodes (Appendix \ref{sec_c}), their weights (Appendix \ref{sec_d}), and the interpolation of measured VDFs on polynomial quadrature grids formed with the nodes (Appendix \ref{sec_e}). This decomposition chain provides the practical route from an observed VDF to a set of HH, HL, or H3 coefficients whose squared amplitudes can be interpreted as velocity-space spectral power.

\begin{figure*}[htb!]
\centering
\includegraphics[width=0.85\textwidth]{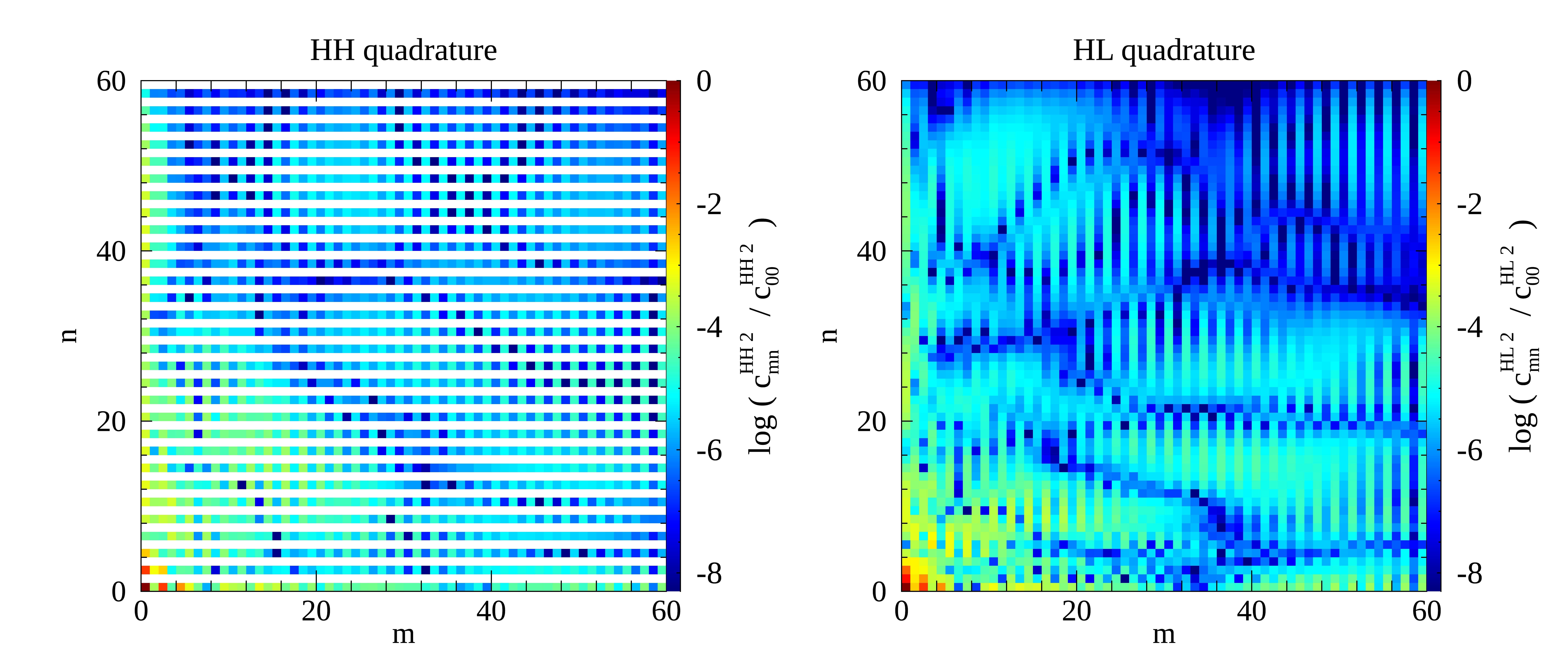}
\caption{2D spectra of the HH and HL polynomials obtained from a SolO/SWA-PAS proton VDF measured on 2022 March 8 at 14:45:22 UT. The spectra are represented by the squared coefficients normalized by their maximum, ${c_{mn}^{HH}}^2 / {c_{00}^{HH}}^2$, estimated from the HH (\textit{left}) and ${c_{mn}^{HL}}^2 / {c_{00}^{HL}}^2$ from HL quadrature grids (\textit{right}), respectively.}\label{fig:1}
\end{figure*}

\subsection{Coefficients of polynomial bases}\label{sec_24}

The polynomial coefficients in equations \ref{eq10} and \ref{eq11} can be estimated using the quadrature rule expressed as equation \ref{eqb1}. For the HH case, equation \ref{eq10} can be converted into a weighted sum as

\begin{equation}
    \begin{split}
    c_{mn}^{HH} = \sum_{ij} f^{HH}(\xi_{\parallel,i},\xi_{\perp,j}) w_{H}(\xi_{\parallel,i}) w_{H}(\xi_{\perp,j}) \times \\ 
    e^{\xi_{\parallel,i}^2} \eta_{m}(\xi_{\parallel,i}) e^{\xi_{\perp,j}^2} \eta_{n}(\xi_{\perp,j}),
    \end{split}
    \label{eq54}
\end{equation}

\noindent where $\xi_{\parallel,i}$ and $\xi_{\perp,j}$ are Hermite nodes, and $w_H$ is the corresponding Hermite node weight. For the HL case, equation \ref{eq11} can be similarly expressed into the weighted sum as

\begin{equation}
    \begin{split}
    c_{mn}^{HL}=\sum_{i,j} f^{HL}(\xi_{\parallel,i},\sqrt{\mu_j}) w_H(\xi_{\parallel,i}) w_L(\mu_j) \times \\
    e^{\xi_{\parallel,i}^2}\eta_m(\xi_{\parallel,i}) e^{\mu_j}\lambda_n^0(\mu_j),
    \end{split}
    \label{eq56}
\end{equation}

\noindent where $\mu_j$ are Laguerre nodes, with $\mu_j=\xi_j^2$, and $w_L$ are the corresponding node weights. 

Figure \ref{fig:1} shows the calculated polynomial coefficients, $c_{mn}^{HH}$ (\textit{left panel}) $c_{mn}^{HL}$ (\textit{right panel}), as a 2D ``map'' of the degree to which a particle distribution deviates from a bi-Maxwellian. Motivated by Parseval's theorem, we consider ${c_{mn}^{HH}}^2$ and ${c_{mn}^{HL}}^2$ normalized by ${c_{00}^{HH}}^2$ and ${c_{00}^{HL}}^2$, respectively, that can be interpreted as the relative power of each velocity-space mode with respect to the lowest-order component. For the HH decomposition shown in the \textit{left panel}, all coefficients of the odd-order Hermite polynomials in the perpendicular direction vanish. This occurs due to the even $f^{HH}(\xi_{\parallel},\xi_{\perp})$ under $f^{HH}(\xi_{\parallel},\xi_{\perp})=f^{HH}(\xi_{\parallel},-\xi_{\perp})$ and due to the parity of the odd-order Hermite polynomial, \textit{i.e.,} $H_m(-x)=(-1)^m H_m(x)$. In the HL representation, as the perpendicular direction is expanded in $\mu$, no analogous even--odd separation appears. Additionally, some organized high-order features are also visible in both HH and HL representations at $m,n\gtrsim10$, which are caused by high-order oscillations. This can result from finite sampling, interpolation onto the quadrature grid, instrumental velocity-space coverage, or genuine small-scale VDF structure; these effects are discussed further in the Appendices from \ref{sec_f} to \ref{sec_i}.

\begin{figure*}[htb!]
\centering
\includegraphics[width=0.9\textwidth,height=0.92\textheight,keepaspectratio]{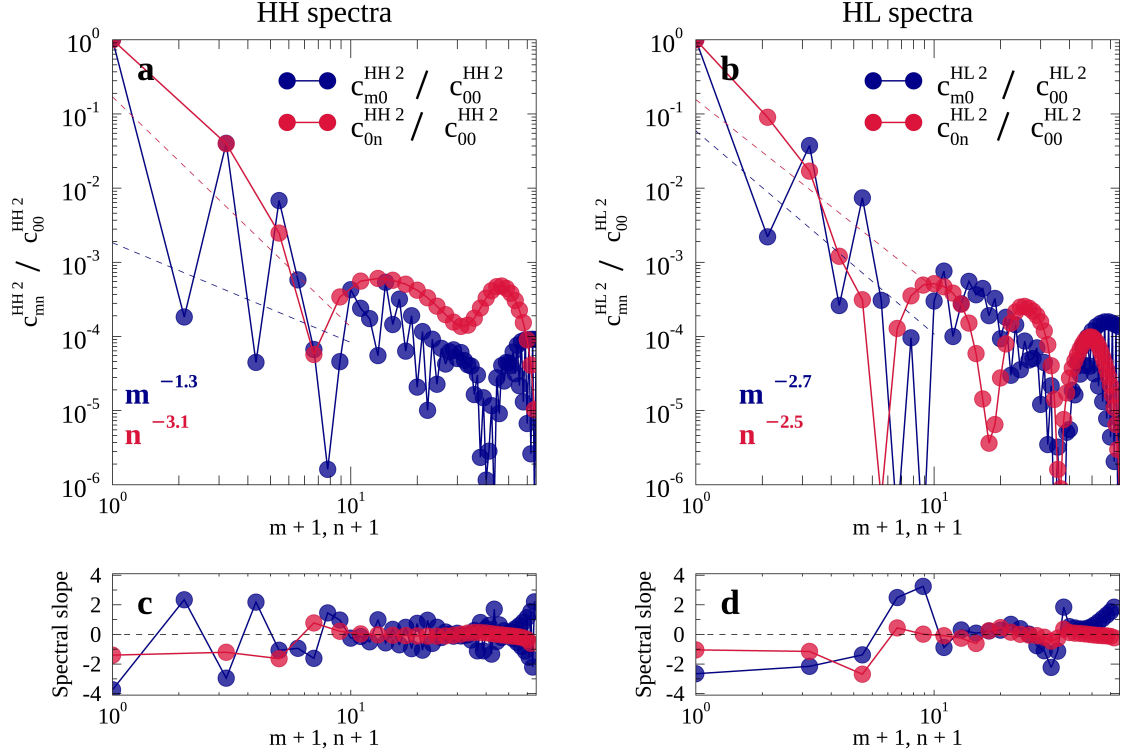}
\caption{Spectra of the quadrature coefficients for the (a) HH and (b) HL decompositions, \textit{i.e.,} ${c_{m0}^{HH}}^2 / {c_{00}^{HH}}^2$ or ${c_{m0}^{HL}}^2 / {c_{00}^{HL}}^2$ (blue solid line) and ${c_{0n}^{HH}}^2 / {c_{00}^{HH}}^2$ or ${c_{0n}^{HL}}^2 / {c_{00}^{HL}}^2$ (red solid line), respectively. The corresponding average slopes (dashed lines) are obtained from linear fits over orders $1 \leq m,n \leq 10$ in log--log space. Only even orders are used for the HH spectrum as the odd-order powers vanish under the assumption of gyrotropy. Panels (c) and (d) show the local spectral slopes for the HH and HL decompositions, respectively, estimated as $\log{({c_{(m+2)0}^{HH}}^2 / {c_{m0}^{HH}}^2})/\log{(1+2/m)}$ or $\log{({c_{(m+2)0}^{HL}}^2 / {c_{m0}^{HL}}^2})/\log{(1+2/m)}$ (blue solid line) and $\log{({c_{0(n+2)}^{HH}}^2 / {c_{0n}^{HH}}^2})/\log{(1+2/n)}$ or $\log{({c_{0(n+2)}^{HL}}^2 / {c_{0n}^{HL}}^2})/\log{(1+2/n)}$ (red solid line).}\label{fig:2}
\end{figure*}

\begin{figure*}[htb!]
\centering
\includegraphics[width=0.99\textwidth,height=0.92\textheight,keepaspectratio]{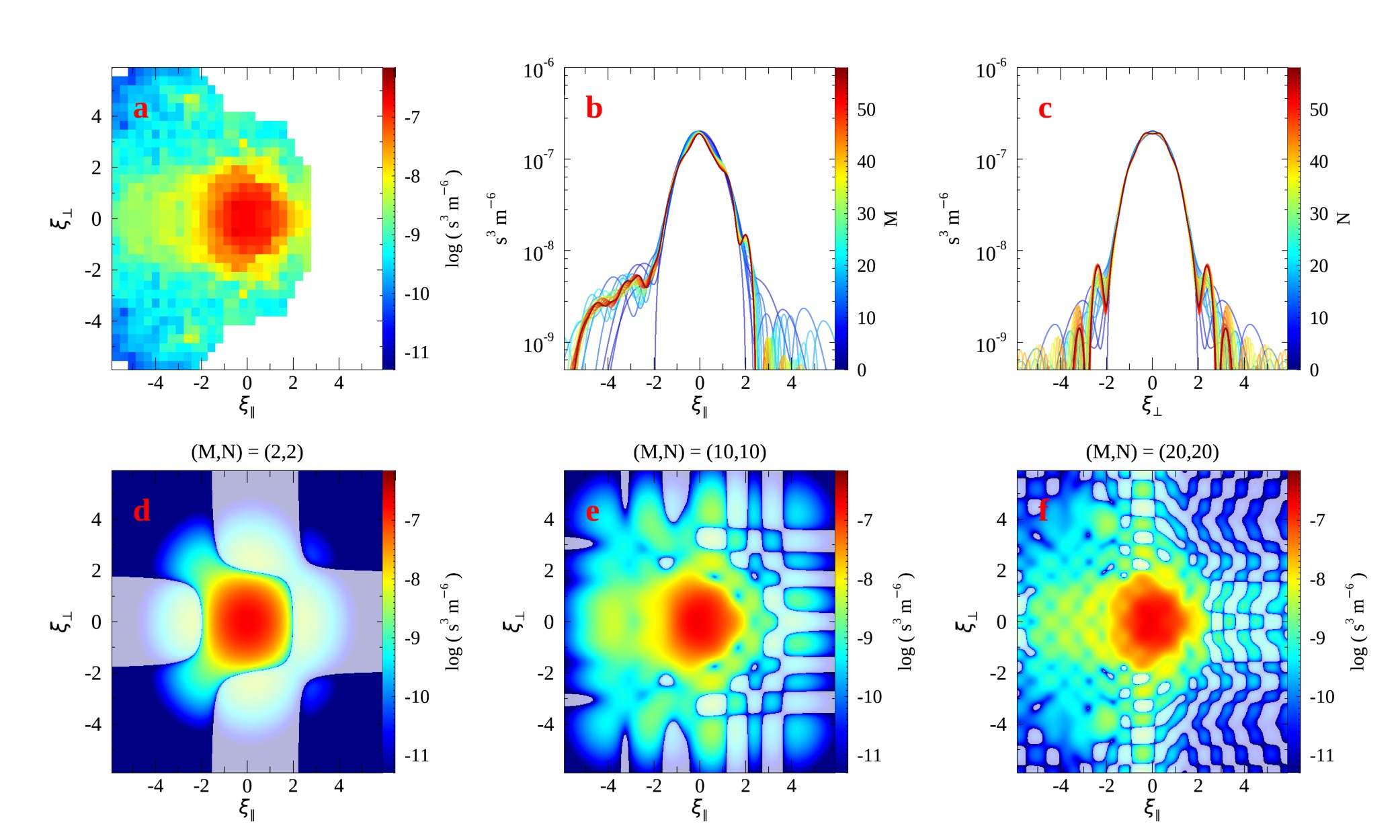}
\caption{Reconstructed VDFs obtained from the HH representation using equation \ref{eq8}. Panel (a) shows the VDF interpolated onto an HH grid for direct comparison with the reconstructed VDFs (see Figure \ref{fig:e1}). Panels (b) and (c) present the parallel ($\xi_{\perp}=0$) and perpendicular ($\xi_{\parallel}=0$) cuts, respectively, of the reconstructed VDFs for orders from 0 to 58. 2D reconstructed VDFs with truncation orders $(M,N)$ of (d) 2, (e) 10, and (f) 20 are also shown. Velocity space is normalized to the proton-core thermal speed. The shaded areas in panels (d), (e), and (f) indicate the negative-valued regions caused by the mathematical representations.}\label{fig:3}
\end{figure*}


\section{Applications of polynomial decomposition}\label{sec_3}

The applications presented in this section can be summarized as a practical workflow for applying polynomial decomposition to spacecraft VDFs. First, the measured VDF should be transformed into a MFA frame, and each velocity component is shifted by the corresponding fitted proton-core drift velocity and scaled by the proton-core thermal speed, as defined in equation \ref{vnml}. Second, the basis should be selected according to the geometry of the problem; HL is preferred for reduced gyrotropic VDFs, and HH is useful for diagnostic comparison. Third, the VDF should be interpolated onto the corresponding quadrature nodes and decomposed into expansion coefficients resulting in spectra. The convergence of the spectra is recommended to be tested against quadrature order. Coefficients above the effective high-order (\textit{e.g.,} $>10$) floor can be excluded to reconstruct a smoother VDF, which may then be used for population fitting, velocity-space-gradient estimation, or comparison of VDF evolution across different plasma environments. The following subsections demonstrate these applications using selected spacecraft observations.

\subsection{Noise reduction and reconstruction of observed VDFs}\label{sec_31}

One practical application of polynomial VDF decomposition is the identification of high-order structure that is likely dominated by noise, providing a useful basis for selecting a truncation range for VDF reconstruction.

Figure \ref{fig:2} presents the one-dimensional (1D) parallel (${c_{m0}^{HH}}^2$ or ${c_{m0}^{HL}}^2$; blue) and perpendicular (${c_{0n}^{HH}}^2$ or ${c_{0n}^{HL}}^2$; red) spectra obtained from the HH (Figure \ref{fig:2}a) and HL (Figure \ref{fig:2}b) decompositions. In both cases, the spectra decrease with increasing polynomial order over the low-order range ($\lesssim 10$) and subsequently exhibit flattening and oscillatory behavior at higher orders. Specifically, in the HH decomposition, the spectral flattening is observed at approximately 8 for both directions, while in the HL decomposition it begins near $m\approx10$ in the parallel direction and $n\approx9$ in the perpendicular direction. The low-order decay is characterized by power-law fits in log--log space over the order range [1,10], which are represented as dashed lines, yielding $m^{-1.3}$ and $n^{-3.1}$ for the HH spectra and $m^{-2.7}$ and $n^{-2.5}$ for the HL spectra.

The variation of local spectral slopes is more apparent in the Figures \ref{fig:2}c and \ref{fig:2}d, where the slopes are calculated as gradients between two points separated by two index units in log--log space, \textit{e.g.,} $\log{({c_{(m+2)0}^{HH}}^2/{c_{m0}^{HH}}^2}) / \log{(1 + 2/m)}$. This trend clearly shows that spectral flattening begins near the order of 10, which can be interpreted as a noise effect.

Using equation \ref{eq8} or \ref{eq9}, the VDF can be reconstructed from the estimated polynomial expansion coefficients, which is the inverse of the decomposing procedure in equations \ref{eq10} and \ref{eq11}. When the polynomial order is truncated to avoid high-order noise effects, the VDF is reconstructed using fewer polynomial components. Figure \ref{fig:3} presents reconstructed VDFs with different truncation orders, together with the interpolated VDF on the HH grid for comparison (Figure \ref{fig:3}a). Reconstructed VDFs with cutoff orders $M,N$ varying from 0 to 58, represented by different colors, where 59 is the maximum order for the $60\times60$ quadrature grid, are shown in the \textit{upper panels}, which present 1D parallel (Figure \ref{fig:3}b) and perpendicular (Figure \ref{fig:3}c) cuts. The recovery of a bi-Maxwellian distribution for $M=N=0$ is confirmed, and truncation orders beyond 10 reconstruct VDFs with significant ripples in the high-speed regimes, \textit{e.g.,} $|\xi_{\parallel}|, \xi_{\perp}>2$. The multiple peaks appearing at high truncation orders should not be interpreted by themselves as distinct proton populations, --- especially when they are not supported by the measured VDF --- instead, they indicate mathematical representations of the observed VDF. \textit{Bottom panels} show 2D reconstructed VDFs with truncation orders of $M=N=$ 2, 10, and 20. The VDF reconstructed with $M=N=2$ (Figure \ref{fig:3}d) does not reproduce the higher-energy regimes because of overtruncation. The VDF reconstructed with $M=N=10$ (Figure \ref{fig:3}e) exhibits a smooth distribution consisting of a zero-centered Maxwellian-like core and a moderate high-energy tail. The VDF reconstructed with $M=N=20$ (Figure \ref{fig:3}f) includes ripples associated with high-order noise in velocity space. The reconstructed VDFs commonly contain negative-valued regions, denoted by the shaded areas, that are not physically interpretable; instead, they are regarded as a part of the mathematical representations. Therefore, the polynomial spectrum obtained from VDF decomposition can effectively identify and mitigate potential noise effects, enabling noise-reduced analysis.

\begin{figure*}[htb!]
\centering
\includegraphics[width=0.99\textwidth,height=0.92\textheight,keepaspectratio]{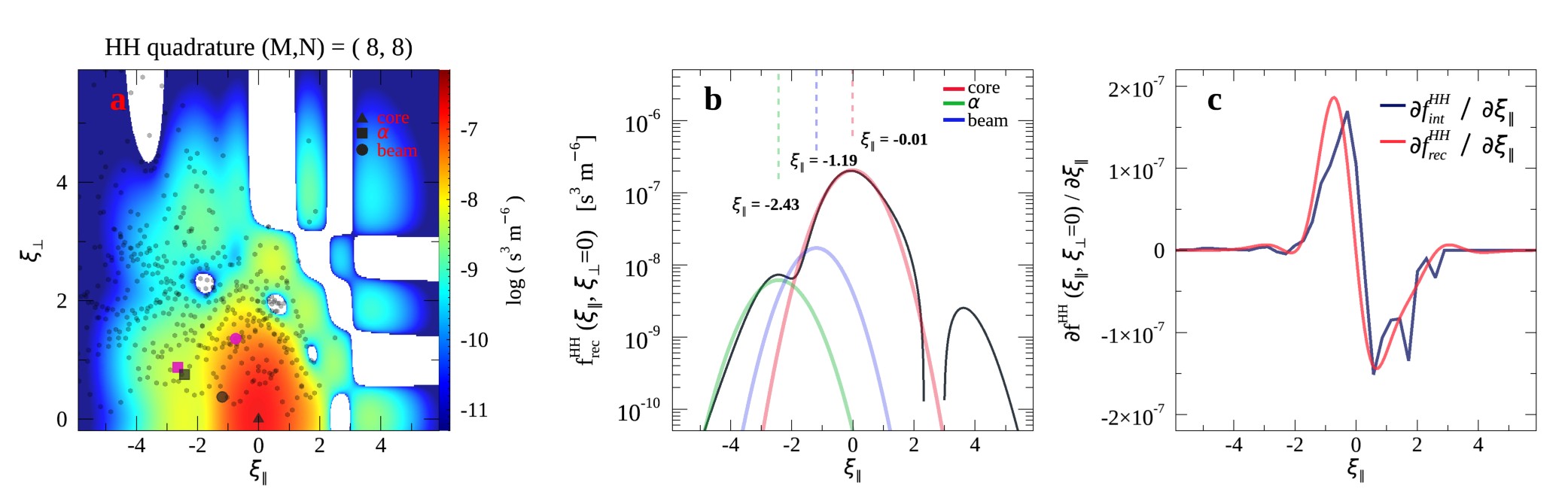}
\caption{Estimations of three ion populations. (a) Reconstructed VDF with Hermite polynomials $m \leq 8$ and $n \leq 8$ (Figure \ref{fig:3}b). The centers of the proton core, beam, and $\alpha$-particle populations are marked by a triangle, circle, and square, respectively. The black dots indicate the $\xi$-space coordinates of SWA-PAS where the VDF is measured (see Figure \ref{fig:e1}a). Note that the marks with magenta (black) color indicate the populations estimated from the observed (reconstructed) VDF. The negative-valued regions are not shown for a clear visualization. (b) Parallel cut ($\xi_{\perp}=0$) of the reconstructed VDF ($f_{\rm rec}^{HH}$; black curve) and the estimated ion populations: core (red), beam (blue), and $\alpha$ particles (green). The locations of their peaks in $\xi_{\parallel}$ are $-0.01$, $-1.19$, and $-2.43$, respectively. (c) Comparison of $\xi_{\parallel}$-gradients of the reconstructed ($\partial f_{\rm rec}^{HH} / \partial \xi_{\parallel}$; red) and interpolated VDF ($\partial f_{\rm int}^{HH} / \partial \xi_{\parallel}$; blue) on to HH quadrature.}\label{fig:4}
\end{figure*}

\begin{figure*}[htb!]
\centering
\includegraphics[width=0.71\textwidth,height=0.92\textheight,keepaspectratio]{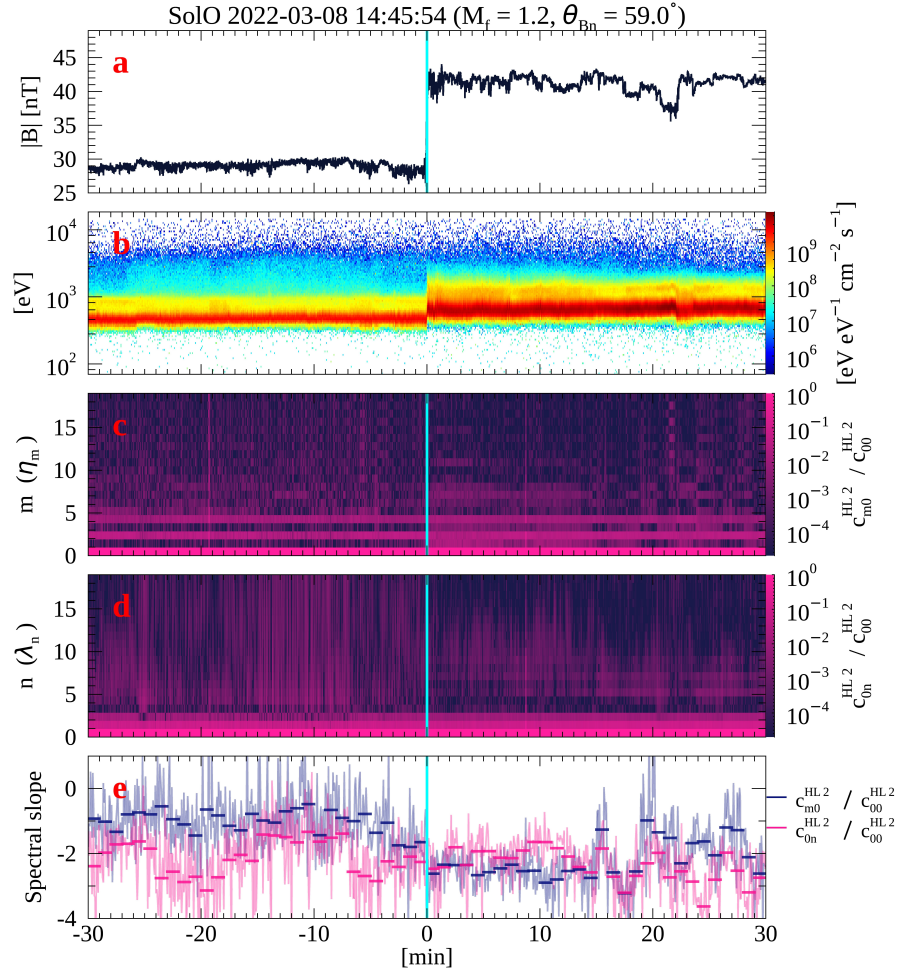}
\caption{Evolution of plasma parameters and VDF spectra across an interplanetary shock observed by SolO at 2022 March 8 14:45:54. Shown are (a) magnetic field, (b) ion differential energy flux, (c) Hermite-polynomial spectrum for the parallel VDF cut, ${c_{m0}^{HL}}^2 / {c_{00}^{HL}}^2$, (d) Laguerre-polynomial spectrum for the perpendicular cut, ${c_{0n}^{HL}}^2 / {c_{00}^{HL}}^2$, and (e) low-order spectral slopes ($1\leq m,n\leq10$) for the parallel-cut spectrum (blue) and perpendicular-cut spectrum (magenta). Horizontal bars denote average slopes within 1 minute sliding windows. The vertical cyan lines indicate the shock crossing, represented as 0.}\label{fig:5}
\end{figure*}

\begin{figure*}[htb!]
\centering
\includegraphics[width=0.71\textwidth,height=0.92\textheight,keepaspectratio]{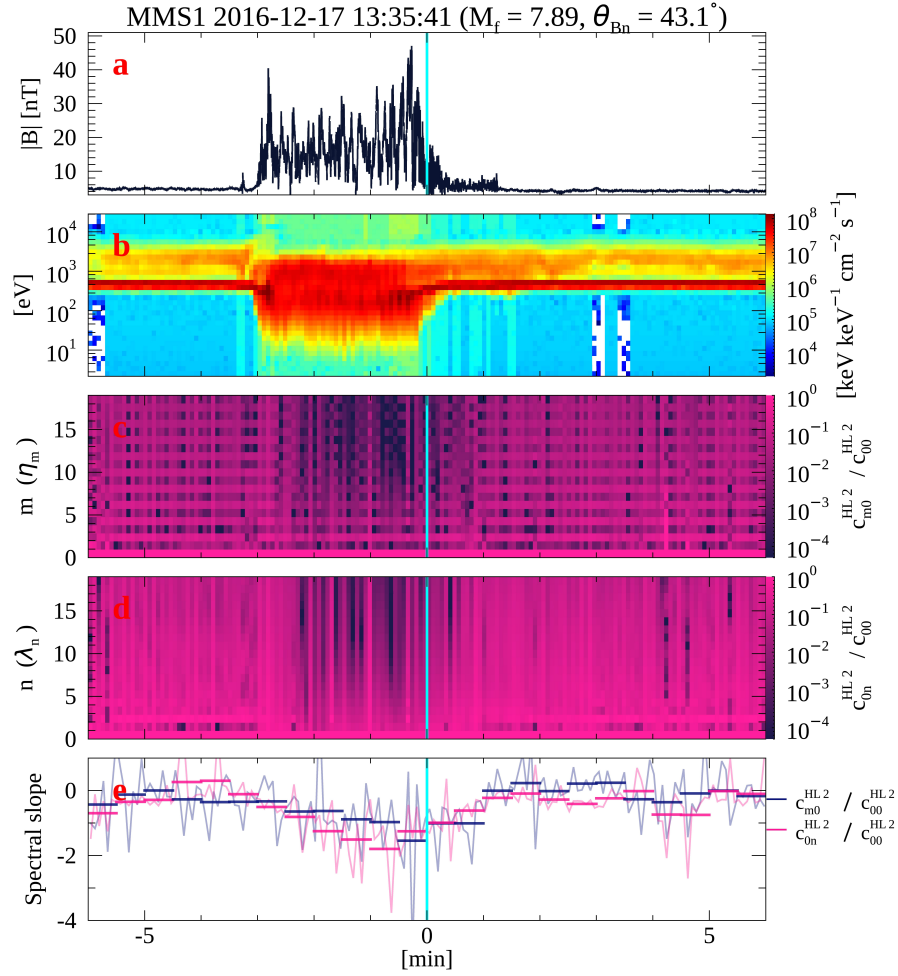}
\caption{Evolution of plasma parameters and VDF spectra across Earth's bow shock observed by MMS1 during 2016 December 17 13:32:43--13:35:41. The figure follows the same format as Figure \ref{fig:5}, except that panel (e) uses 30 s averaging windows.}\label{fig:6}
\end{figure*}

\begin{figure*}[htb!]
\centering
\includegraphics[width=0.98\textwidth,height=0.92\textheight,keepaspectratio]{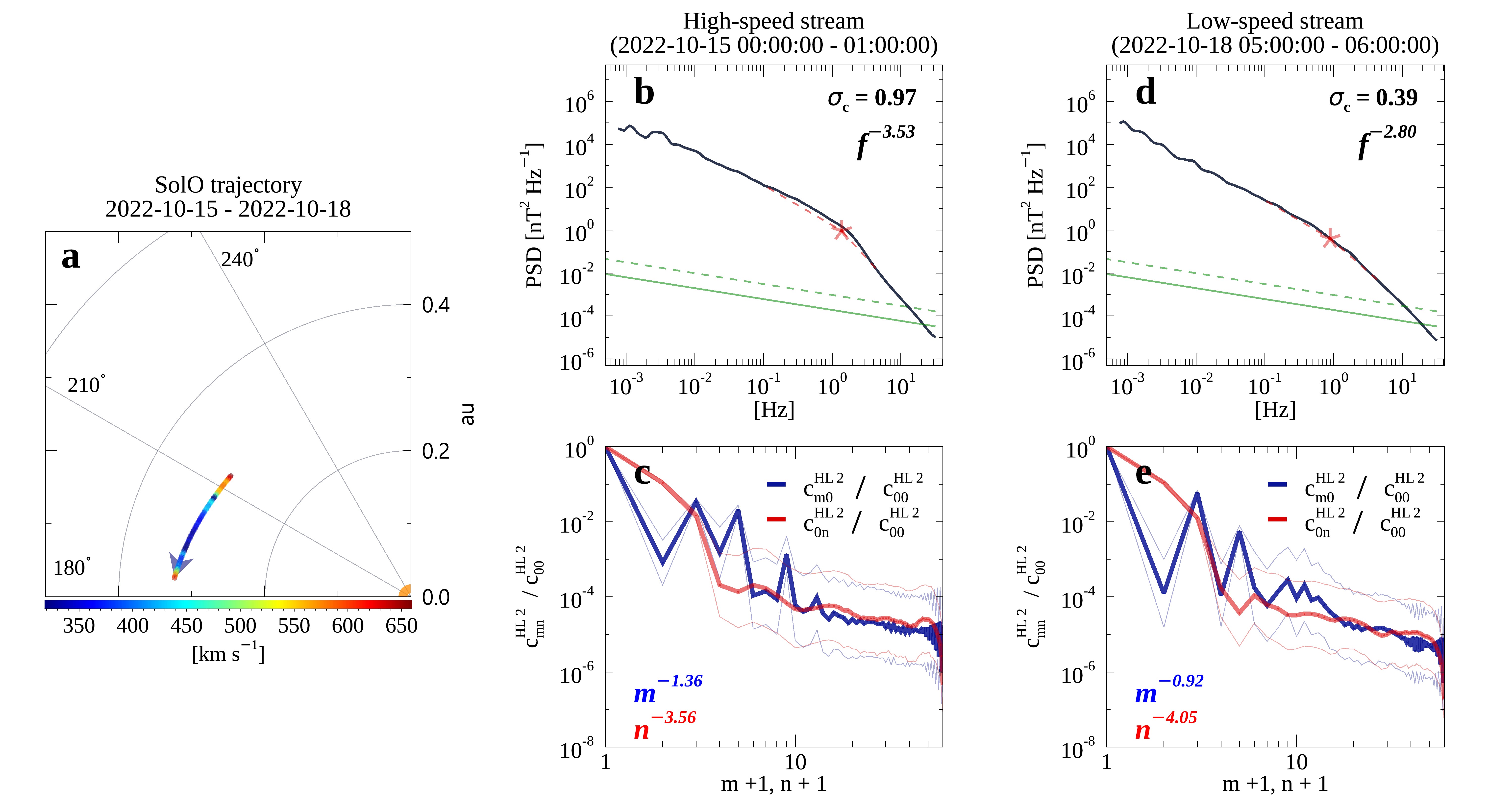}
\caption{Magnetic field and VDF spectra estimated in turbulent solar wind streams with different cross helicities, $\sigma_{c}$. (a) Trajectory of SolO from 2022 October 15 00:00:00 to 2022 October 18 23:59:59 in the ecliptic plane of the Carrington frame, where color indicates stream speed. (b) Magnetic-field spectrum measured by the burst-mode MAG instrument. The spectral break and fitted two-segment piecewise linear function \citep{Park2023} are marked by the red asterisk and dashed line, respectively. The MAG noise floor is denoted by the green solid line, and the cutoff noise floor used for estimating the transition-range spectral slope, defined as five times the MAG noise floor, is denoted by the green dashed line \citep{park2025}. (c) Averaged VDF spectra for the parallel (${c_{m0}^{HL}}^2/{c_{00}^{HL}}^2$; blue) and perpendicular (${c_{0n}^{HL}}^2/{c_{00}^{HL}}^2$; magenta) cuts during the high-speed stream from 2022 October 15 00:00:00 to 01:00:00. The thinner lines show the standard deviation of the corresponding average spectrum. Panels (d) and (e) present the magnetic-field and VDF spectra, respectively, in the same format as panels (b) and (c), but for the low-speed stream from 2022 October 18 05:00:00 to 06:00:00. Marked power laws of polynomial spectra in panels (c) and (e) are estimated by linear fitting in log--log space over $1\leq m,n\leq10$.}\label{fig:7}
\end{figure*}

A reliable VDF with reduced noise enables several applications. For example, robust fits of analytical models to noise-reduced VDFs can be considered. In Figure \ref{fig:4}, the ion populations are estimated by fitting bi-Maxwellian components to the reconstructed VDF using a Levenberg--Marquardt least-squares algorithm that minimizes $\chi^{2}$. The fitting is initialized in three velocity-space regions. The proton core is initialized around the origin of the normalized velocity space as described in Section \ref{sec_21}. The proton beam is initialized in semi-annular regions, defined here as the portion of the reduced $(\xi_{\parallel},\xi_{\perp})$ plane bounded by two radial distances, $0.1v_{\mathrm A}/w_{\parallel}$ and $1.1v_{\mathrm A}/w_{\parallel}$; thus, $0.1v_{\mathrm A}/w_{\parallel}\leq |\xi| \leq1.1v_{\mathrm A}/w_{\parallel}$, where $|\xi|=(\xi_{\parallel}^{2}+\xi_{\perp}^{2})^{1/2}$, $v_{\mathrm{A}} = \mathbf{|B|} / \sqrt{\mu_{0}\rho_{p}}$, and $\rho_{p}$ is the proton mass density. The $\alpha$ population is initialized within the expected mass-per-charge shifted region, $[(\sqrt{2}-1)v_{\parallel}-0.3v_{\mathrm A}]/w_{\parallel}\leq |\xi| \leq[(\sqrt{2}-1)v_{\parallel}+v_{\mathrm A}]/w_{\parallel}$.

Figure \ref{fig:4}a compares the fitted population centers --- proton core, proton beam, and $\alpha$ particles --- obtained from the observed VDF and from the reconstructed VDF with $M=N=8$. The core and $\alpha$-particle locations are relatively stable, whereas the beam center shows some difference between the two fits, indicating the effect of noise-filtering and/or the sensitivity of the secondary-population identification to finite-order reconstruction. Nevertheless, the reconstructed VDF provides a smooth, noise-reduced representation of the large-scale velocity-space morphology, which can facilitate population fitting and provide more stable inputs for subsequent analyses. The corresponding parallel cut at $\xi_{\perp}=0$ is shown in Figure \ref{fig:4}b. Since the bi-Maxwellian fits are performed in 2D velocity space, the fitted population centers do not necessarily coincide with the maxima of the 1D parallel cut.

Figure \ref{fig:4}c compares the $\xi_{\parallel}$-gradients of the reconstructed VDF, $f_{\rm rec}^{HH}$, and the interpolated VDF, $f_{\rm int}^{HH}$, on the HH quadrature grid. The reconstructed VDF provides a smoother gradient while preserving the dominant large-scale velocity-space structure, which may support subsequent kinetic analyses such as the estimation of instability growth rates \citep{coburn2024}.

\subsection{Evolution of VDFs across shocks in space}\label{sec_32}

Collisionless shocks produce substantial modifications of particle VDFs, including heating, reflected populations, and nonthermal structure. We apply the polynomial-decomposition method to two illustrative events: a weak quasi-perpendicular interplanetary shock observed by SolO and a stronger, more quasi-parallel terrestrial bow-shock crossing observed by MMS1. As the two events differ in shock environment and geometry, differences in their spectral evolution can be interpreted as event-specific examples of VDF modification rather than as statistical conclusions regarding shock type or shock geometry.

Figure \ref{fig:5} presents the evolution of plasma parameters, including (a) magnetic field and (b) omnidirectional ion differential energy flux, and spectral properties, including (c) ${c_{m0}^{HL}}^2/{c_{00}^{HL}}^2$, (d) ${c_{0n}^{HL}}^2/{c_{00}^{HL}}^2$, and (e) low-order spectral indices estimated as linear fits over the order 1 to 10, across an interplanetary shock observed by SolO at 2022 March 8 14:45:54, as reported by \citet{trotta2024}. This event is a weak quasi-perpendicular --- the angle between the background magnetic field and the shock normal $\theta_{\rm Bn}=59^{\circ}$ --- shock with fast-mode Mach number $M_{\rm f}=1.2$. Although no distinct field aligned beams reflected from the shock front are evident within our analysis intervals, wave structures and superthermal foreshock ions are present in the magnetic field and energy flux profiles (Figures \ref{fig:5}a and \ref{fig:5}b) as reported in \citep{trotta2024}.

The parallel spectral power, ${c_{m0}^{HL}}^2/{c_{00}^{HL}}^2$, increases downstream of the shock, from 0 to 15 minutes after the shock crossing, over orders approximately between 1 and 8, whereas no significant power variation is observed in the upstream. The perpendicular spectral power, ${c_{0n}^{HL}}^2/{c_{00}^{HL}}^2$ (Figure \ref{fig:5}d), forms enhanced regions upstream from $-26$ to $-25$ minutes and from $-15$ to $-7$ minutes, where the power is amplified up to order $\sim20$. This enhancement may be associated with a modest increase in ion flux at 1--3 keV between $-26$ to $-7$ minutes shown in Figure \ref{fig:5}b. The physical mechanism responsible for the relatively low ${c_{0n}^{HL}}^2/{c_{00}^{HL}}^2$ between $-25$ and $-15$ minutes has not yet been identified. Additionally, from approximately $-7$ minutes onward, a reduction in ${c_{0n}^{HL}}^2/{c_{00}^{HL}}^2$ is observed mainly at orders $3\lesssim n\lesssim5$. This spectral variation is reflected in the evolution of the low-order spectral indices (Figure \ref{fig:5}e). The parallel spectral slope (blue) remains near $-1$ upstream and steepens toward $-2$ near the shock front, whereas the enhanced upstream regions of ${c_{0n}^{HL}}^2/{c_{00}^{HL}}^2$ have shallower indices, approximately $-1.5$, that steepen to approximately $-2.5$ beginning about 7 minutes before shock passage. The concurrent spectral steepening in both directions beginning near $-7$ minutes may be related to a slight increase in the energy flux near $0.9$ keV, where $\alpha$-particles are likely observed. The correlation of $\alpha$-particle abundance with spectral-power variations will be examined in a separate study. Although beam-like reflected ions within the foreshock region are not clearly observed, the distributions are strongly deformed near and across the shock.

Steepening of the spectral slopes suggests a reduction of fine structures in velocity space and a redistribution of spectral power toward lower-order modes associated with the proton core thermalization. Another example of spectral evolution is presented in Figure \ref{fig:6}, using an Earth's bow shock crossing observed by MMS1 at 2016 December 17 13:35:41 \footnote[6]{https://sharp.fmi.fi/shock-database/}. This event is a strong quasi-parallel shock with $M_{\rm f}\approx7.9$ and $\theta_{\rm Bn}\approx43^{\circ}$. Ion foreshock particles are observed throughout the upstream interval, particularly in the energy range of approximately $0.8$--$4~\mathrm{keV}$, as indicated by ion differential energy-flux profiles (Figure \ref{fig:6}b), while the magnetic field profile (Figure \ref{fig:6}a) exhibits potential wave activities extending up to $1.5$ minutes from the shock crossing. Although isolated impacts from this ion foreshock on the spectra cannot be determined from the event alone, an accompanying reduction of high-order spectral power with a localized enhancement of low-energy particle flux in the $60$--$200~\mathrm{eV}$ range is observed in both the parallel and perpendicular directions (Figures \ref{fig:6}c and \ref{fig:6}d). The low-order spectral indices (Figure \ref{fig:6}e) also steepen across the shock, varying from $-0.5$ to $-1.5$ in the parallel direction and from approximately $-0.2$ to $-2$ in the perpendicular direction. Together, the two events demonstrate that the distribution and slopes of expansion-coefficient power vary systematically across individual shock crossings.

\subsection{Variation of VDF in turbulence}\label{sec_33}

Turbulent solar-wind streams with different Alfv{\'e}nic properties can exhibit distinct ion-scale (\textit{e.g.,} ion gyroradius) dissipation signatures often observed in magnetic field spectra; the Alfv{\'e}nicity may control these signatures by regulating the transfer of turbulent energy toward sub-ion scales, which is understood in the framework of `helicity barrier' \citep{meyrand2021,squire2022,bowen2024,mcintyre2025,panchal2025}. We compare a high-speed and a low-speed stream observed by SolO and characterize their Alfv{\'e}nicity using the normalized cross helicity, $\sigma_{\rm c}=\frac{2\langle \delta\mathbf{v}\cdot\delta\mathbf{b}\rangle}{\langle |\delta\mathbf{v}|^2\rangle+\langle |\delta\mathbf{b}|^2\rangle}$, which measures the degree of Alfv{\'e}nic correlation between the velocity fluctuation $\delta\mathbf{v}$ and the magnetic-field fluctuation $\delta\mathbf{b}$ expressed in Alfv{\'e}n-speed ($v_{\mathrm A}$) units. The comparison examines how differences in $\sigma_{\rm c}$ and in the near-ion-scale (or transition range) magnetic field spectrum are accompanied by changes in the parallel and perpendicular spectra of expansion-coefficient power.

Figure \ref{fig:7} presents magnetic-field and VDF spectra in different turbulent streams observed by SolO from 2022 October 15 00:00:00 to 2022 October 18 23:59:59. As indicated by different colors in Figure \ref{fig:7}a, the average speeds of the two stream intervals are approximately 650 $\rm km \ s^{-1}$ and 350 $\rm km \ s^{-1}$, denoted as high- and low-speed streams, respectively. The high-speed stream (Figure \ref{fig:7}b) has higher $\sigma_{\rm c}$ ($\approx0.97$), exhibiting a steeper transition-range magnetic-field spectral index of $-3.53$, whereas the low-speed stream (Figure \ref{fig:7}d) contains $\sigma_{\rm c}\approx0.39$ and a shallower index of $-2.80$. The average low-order VDF spectral slopes in the parallel (blue) and perpendicular (red) directions during the high-speed stream (Figure \ref{fig:7}c) are $-1.36$ and $-3.56$, respectively, whereas those during the low-speed stream (Figure \ref{fig:7}e) are $-0.92$ and $-4.05$, respectively.

The two intervals exhibit anisotropic VDF spectra. Compared to the low-speed stream, the high-speed stream shows a steeper parallel spectral slope but a flatter perpendicular spectral slope, with a steeper transition-range magnetic-field spectrum. The correspondence between these VDF spectral differences and the magnetic-field turbulence suggests that the decomposition method may provide a useful diagnostic of velocity-space signatures associated with wave--particle energy transfer. This possible connection is discussed further in Section \ref{sec_4}.


\section{Discussion \label{sec_4}}

We present a framework for decomposing solar wind VDFs using orthogonal polynomial bases. This method converts the shape of a measured VDF into a set of spectral coefficients, enabling deviations from a Maxwellian core to be quantified in a model-independent manner. In analogy with Fourier spectra of electromagnetic fluctuations, the polynomial spectrum provides information on how VDF power is distributed across velocity-space scales and offers a compact description of VDF morphology such as smooth Maxwellian-based structures, nonthermal features, and velocity-space gradients.

\subsection{Physical interpretation of polynomial order and velocity-space structure}\label{sec_41}

The physical meaning of polynomial order follows from the oscillatory nature of the basis functions. Higher polynomial orders therefore represent progressively finer velocity-space structure. In Fourier analysis, the physical scales correspond to the frequencies or wavenumbers of the Fourier basis functions. Similarly, the polynomial order controls an effective ``wavelength'' in velocity space and can capture velocity-space structures at corresponding scales. For example, low orders describe the Maxwellian core and its low-moment-like deviations, such as drift, temperature anisotropy, and skewness, whereas high orders represent sharply deformed structures, including beams, deficits, reflected particles, shell-like distributions, and phase-space ripples.

This interpretation is closely related to previous studies of velocity-space ``cascades''. MMS1 observations and kinetic simulations have shown that structured VDFs can exhibit extended Hermite spectra, suggesting a transfer of free energy toward small velocity-space scales \citep{servidio2017,pezzi2018,cerri2018}. Similar interpretations have been applied to near-Sun ion VDFs observed by PSP \citep{larosa2025}. 

From a kinetic perspective, small-scale velocity-space structure can be generated by several processes. Linear phase mixing associated with Landau damping produces fine structure primarily in $v_\parallel$ and transfers perturbation energy from low to high Hermite moments, although nonlinear plasma echoes and turbulent advection may partially suppress this Hermite-space flux \citep{schekochihin2016,meyrand2019}. Direct field-particle correlation measurements have identified velocity-space signatures of electron Landau damping in magnetosheath turbulence, demonstrating that resonant energy transfer can be localized in velocity space \citep{chen2019,horvath2020}. In contrast, cyclotron interactions, pitch-angle scattering, stochastic heating, and reflected-particle dynamics can generate perpendicular or mixed parallel--perpendicular gradients, which should appear as enhanced perpendicular Hermite/Laguerre power or diagonal structures in 2D coefficient maps \citep{afshari2024,coburn2024,nastac2024}. Therefore, high-order coefficient power may contain physical information on kinetic energization, provided that it can be distinguished from numerical and instrumental artifacts.

One possible example of such wave--particle coupling, specifically involving ion cyclotron waves (ICWs), is presented in Section \ref{sec_33}. In the helicity-barrier framework, strongly imbalanced Alfvénic turbulence, characterized by large $\sigma_{\rm c}$, can inhibit the forward cascade of turbulent energy to sub-ion scales and generate ICWs near ion scales \citep{bowen2024}. These waves can subsequently transfer energy to ions through cyclotron-resonant interactions, preferentially causing perpendicular heating. Consistent with this picture, the high-speed stream exhibits a shallower perpendicular VDF spectrum as shown in (Figure \ref{fig:7}c) than the low-speed stream (Figure \ref{fig:7}e), together with a steeper parallel spectrum, which indicates a relative enhancement of perpendicular velocity-space structure. This suggests that VDF spectral slopes may provide diagnostics complementary to magnetic-field spectra and may help connect velocity-space structure with kinetic-scale dissipation processes. The relation among $\sigma_{\rm c}$, the transition-range spectral slope of magnetic-field fluctuations, and the perpendicular VDF spectral steepness is suggestive, but should not yet be interpreted as evidence for a specific dissipation channel. The coefficient-power spectra alone cannot uniquely distinguish ICW-mediated dissipation from other processes capable of generating similar anisotropic VDF structure. Establishing such a connection therefore requires a statistical survey combined with independent identification of the relevant wave modes and resonant particle signatures.

In this context, the different parallel and perpendicular slopes observed in the HH spectra (Figure \ref{fig:2}a) may reflect anisotropic velocity-space structure associated with the aforementioned kinetic processes. However, the HL spectra (Figure \ref{fig:2}b) exhibit power laws distinct from those of the HH spectra and do not show the same anisotropy. This difference cannot be attributed solely to the plasma because the HL expansion is performed in $\xi_{\perp}^{2}$ space and employs a different weighting function and quadrature grid from the HH representation. Consequently, the absolute coefficient powers and spectral slopes obtained from the two bases are not directly equivalent. Therefore, the contrast between the HH and HL spectra should be interpreted within their respective coordinate systems and basis functions, according to how each representation distributes velocity-space structure among polynomial orders and directions.

\subsection{Quantifying VDF alterations at Collisionless Shocks}\label{sec_42}

Collisionless shocks in space provide a useful application of polynomial VDF decomposition as they lead to substantial departures from quasi-equilibrium distributions through particle heating, reflection, scattering, and acceleration \citep{guo2021,johlander2023,trattner2023}. This near-shock physics is reflected in variations of VDFs, including shock-driven flat-top electron distributions \citep{feldman1983,wilson2020,johlander2023,stasiewicz2024}. Although parametric models are effective for well-separated populations, they can become restrictive when multiple components --- such as reflected and/or diffuse particle populations --- overlap, exhibiting broad and irregular morphologies \citep{bonifazi1981,gedalin2022,dimmock2023}.

The examples presented in Section \ref{sec_32} indicate that spectra of expansion-coefficient power can quantify shock-induced VDF modifications. For instance, core broadening and temperature anisotropy predominantly affect lower-order coefficients, while beams, reflected particles, and diffuse populations may distribute power over a broader range of orders. Changes in the corresponding spectral slopes therefore provide a compact measure of the increased complexity and directional structuring of the VDF across shocks, which can diagnose shock-associated physics such as particle interactions with shock-driven instabilities.

\subsection{Practical upper limit of polynomial order}\label{sec_43}

Although high-order polynomials can represent micro- or kinetic-scale structures in velocity space, the polynomial order cannot be increased indefinitely, and its practical upper limit is not yet fully understood. Mathematical studies of polynomial approximation from discrete samples suggest that the highest polynomial degree recoverable stably depends on both the number and distribution of the sampling points \citep{platte2011,adcock2019}. These results do not impose an intrinsic upper limit on the quadrature rule, but imply that the practically useful order in a measurement-based VDF decomposition is data dependent.

Analogous to the Nyquist frequency in Fourier analysis, which corresponds to half the sampling frequency and sets a lower resolvable scale, the relevant limit here is determined by the physical measurement resolution, such as the angular and energy-bin scales of the instruments. As discussed in Appendix \ref{sec_i}, the maximum quadrature order can be estimated by comparing the velocity-space wavelength of each order with the instrumental scales. The upper limit of the quadrature order is much higher than the value of 60 adopted throughout this study. However, as shown in Figures \ref{fig:1} and \ref{fig:2}, noise are still observed at high orders. Thus, this limit should be interpreted as a physical-resolution limit rather than as a quadrature limit. 

As plasma instruments measure VDFs in finite energy-angle bins, the velocity-space resolution depends on both speed and direction. In particular, the angular contribution to the Cartesian velocity-space bin size increases with particle speed. Interpolation from the instrumental grid to the polynomial quadrature grid compensates for coordinate mismatch, but it cannot create information below the original measurement resolution. Consequently, coefficients at orders for which the basis scale is smaller than the instrumental scale are likely dominated by interpolation effects, or sparse sampling rather than by independent plasma structures driven by physical processes such as wave damping and stochastic heating.

This distinction is important for interpreting the high-order flattening observed in polynomial spectra. As shown in Figures \ref{fig:1} and \ref{fig:2}, the spectra decrease nearly exponentially with order and then flatten with oscillations at high order. Similar behavior is reported by \citet{coburn2024} and \citet{larosa2025}, who interpret flattening of the Hermite and Laguerre spectra above order $\sim14$ as a characteristic noise floor. Motivated by the expected spectral flattening caused by noise, analogous to that observed in magnetic-field spectra \citep{woodham2018}, we analyze multiple Hermite and Laguerre spectra during a 6 hr quiet solar wind interval beginning at 2022 July 12 00:00:00 in Appendix \ref{sec_f}. However, we do not identify a typical noise floor in the polynomial spectra (see Figure \ref{fig:f1}), although spectral flattening is observed in the Hermite spectra near $m \approx 10$. Furthermore, in the PSP--SolO comparison shown in Figure \ref{fig:f2}, intervals that may correspond to the same solar wind plasma do not exhibit a common high-order spectral floor. This suggests that high-order behavior is not a universal property of the plasma alone, nor a simple detector noise floor. Instead, it depends on the full measurement--decomposition chain, including instrumental energy-angle sampling, field-of-view coverage, interpolation onto the polynomial quadrature grid, and basis choice. This result further indicates that high-order flattening should be used primarily as a practical truncation guide. Only coefficient power that remains stable under basis changes and quadrature-order variation should be interpreted as physically meaningful velocity-space structure.

\subsection{Selection of polynomial basis}\label{sec_44}

No clear standard has yet been established for choosing between HH and HL decompositions. Considering the gyrotropy assumption, the HL decomposition is generally more natural as its cylindrical geometry avoids the artificial mirroring of perpendicular velocities required in the HH representation. Presented in Section \ref{sec_23}, H3 retains the two perpendicular directions independently and is therefore better suited to resolving nongyrotropic structure. However, the H3 decomposition requires adequate angular coverage and resolution, as discussed in Appendix \ref{sec_h}, even though it can reduce some artificial high-order structure associated with the projection of the measured VDF onto a 2D gyrotropic grid. The preferred representation should be selected according to the expected degree of gyrotropy and the effective sampling quality instead of dimensionality alone.

\subsection{Caveats of VDF reconstruction}\label{sec_45}

Reconstruction of a VDF with a finite polynomial series inevitably produces ripple-like patterns and unphysical negative-valued regions, as shown in Figures \ref{fig:3} and \ref{fig:4}. This behavior indicates an incomplete representation of the actual VDF shape by the finite basis. Although such artifacts are expected to be more pronounced in regions not supported by directly measured data points, analogous to the Gibbs phenomenon that produces over- and undershoots near discontinuities in signals reconstructed with finite Fourier series, artificially filled regions and negative-valued spaces can still appear even after intentional removal of high-order coefficients. Appendix \ref{sec_g} further demonstrates the relation between these regions and the presence of data points, showing that full-sky electron VDF measurements from SWA-EAS still yield reconstructed VDFs with negative-valued regions. This implies that these artifacts are not necessarily caused by the absence of measured particle populations; they can arise from the finite polynomial basis itself and from incomplete recovery of the VDF in regions that are not well constrained by measurements.

This caveat is especially important for applications that require smooth velocity-space gradients. The reconstructed VDF should be regarded as a useful approximation rather than a unique recovery of the true continuous VDF, since gradients of measured distributions are generally noisy, and alternatives are therefore often considered \citep{chen2019}. For example, \citet{coburn2024} obtained smooth gradients from electron VDFs reconstructed with truncated Hermite--Laguerre expansions and used them to estimate instability growth rates. Reconstructed gradients may also be supplied to the Arbitrary Linear Plasma Solver (ALPS; \citealt{verscharen2018,klein2025}) to calculate plasma susceptibilities and dispersion relations from measured VDFs. Nevertheless, the use of VDF reconstruction for applications --- such as instability analysis, fitting with parametric models (Figure \ref{fig:4}), and field-particle energy-transfer diagnostics --- requires additional caution.


\section{Summary and conclusion}\label{sec_5}

The summary of this study are as follows.
\begin{enumerate}
    \item Orthogonal polynomial decomposition provides a nonparametric representation of solar wind VDFs and quantifies deviations from Maxwellian-like equilibrium through spectra of the expansion coefficient power.

    \item For reduced gyrotropic VDFs, the HL basis is generally preferable because it is consistent with the cylindrical geometry of $(\xi_\parallel,\xi_\perp)$ space. The HH basis remains useful for comparison and diagnostics. However, its perpendicular spectrum can be affected by applying an Hermite basis to the mirrored perpendicular coordinate.

    \item Polynomial order has a velocity-space scale interpretation. Higher Hermite or Laguerre orders resolve progressively finer structures in velocity space, which appear as enhanced high-order spectral power and shallower spectral indices. Such high-order power may be associated with several physical mechanisms, including resonant wave-particle interactions through Landau and cyclotron resonances, linear or nonlinear phase mixing, beam and deficit structures, reflected particles, and other kinetic processes.

    \item High-order spectral flattening or oscillatory behavior arise from instrumental noise, partial field-of-view coverage, interpolation across mismatched coordinate systems, finite sampling, and basis choice, as well as from genuine kinetic velocity-space structure.

    \item By truncating coefficients above an effective high-order limit, a noise-reduced and reliable VDF can be reconstructed for population fitting and velocity-space gradient estimation. However, Gibbs-like artifacts and incomplete recovery of the original VDF must be taken into account.

    \item The method is promising for statistical studies of VDF modifications associated with shocks, turbulence, and wave-particle interactions. Its strongest application will come from combining polynomial spectra with physical diagnostics, \textit{e.g.,} quasilinear theory, and linear dispersion solvers such as ALPS.
\end{enumerate}

The polynomial expansion approach therefore provides a useful bridge between measured particle distributions and kinetic plasma physics. It offers a quantitative and reproducible coordinate system in which velocity-space structure, non-Maxwellian features, and associated kinetic processes can be compared across instruments and heliospheric environments. In the longer term, combining polynomial spectra with complementary tools such as linear dispersion solvers, quasilinear diffusion theory, and field--particle correlations may provide a systematic way to connect observed velocity-space structure with specific kinetic processes, including phase mixing, resonant wave--particle interactions, and stochastic heating. Thus, the present framework is not intended only as a mathematical decomposition technique, but as a practical diagnostic basis for future statistical studies of kinetic energy conversion in collisionless space plasmas.


\begin{acknowledgments}

The authors acknowledge the Solar Orbiter SWA and MAG teams, the Parker Solar Probe SWEAP and FIELDS teams, and the MMS1 FPI and FIELDS/FGM teams for the plasma and magnetic-field data used in this study. We acknowledge the use of data available from the Solar Orbiter Archive, the Coordinated Data Analysis Web (CDAWeb), Parker Solar Probe SWEAP and FIELDS Data Archives, and the MMS Science Data Center. We also acknowledge the use of the SHocks: structure, AcceleRation, dissiPation (SHARP) shock database for identifying the MMS1 bow-shock interval analyzed in this study. This work is supported by Estancias Posdoctorales por M\'{e}xico Convocatoria 2025 of La Secretar\'{i}a de Ciencia, Humanidades, Tecnolog\'{i}a e Innovaci\'{o}n (SECIHTI) under CVU 2164068. P.K.'s work is funded by the PAPIIT-DGAPA through the project grant IN100424. D.V. is supported from STFC Consolidated Grant ST/W001004/1. X.BC. thanks PAPIIT DGAPA grant IN106724 and SECIHTI CBF-2023-2024-852 grant. J.S., Z.N., A.P., T.D., and S.S. are supported by the Ministry of Education, Youth, and Sports grant, under contract LUAUS25060. J.T.C acknowledges the financial support provided by the Paris region DIM ORIGINE under grant number IDF-DIM-ORIGINES-2025-2-01.

\end{acknowledgments}

\begin{appendix}

\section{Mathematical expectation of the VDF cascade}\label{sec_a}

We illustrate how deviations from a Gaussian distribution are represented by Hermite polynomials. Given a Gaussian distribution with displacement $\Delta$, the distribution can be written as

\begin{equation}
    f(x) = \frac{1}{\sqrt{\pi}} e^{-(\xi - \Delta)^2}.
    \label{eqa1}
\end{equation}

\noindent Equivalently,

\begin{equation}
    f(x) = \frac{1}{\sqrt{\pi}} e^{-\xi^2} e^{2\xi \Delta - \Delta^2}.
    \label{eqa2}
\end{equation}

\noindent This expression can be interpreted as a Gaussian distribution multiplied by a deviation factor, which is the generating function of the Hermite polynomials,

\begin{equation}
    e^{2\xi \Delta - \Delta^2} = \sum_{m=0}^{\infty} \frac{H_m(\xi)}{m!} \Delta^m.
    \label{eqa3}
\end{equation}

\noindent Thus, 

\begin{equation}
    f(x) = \frac{1}{\sqrt{\pi}} e^{-\xi^2} \sum_{m=0}^{\infty} \frac{H_m(\xi)}{m!} \Delta^m,
    \label{eqa4}
\end{equation}

\noindent which shows that $\Delta\neq0$, corresponding to a non-Gaussian displacement, generates an infinite Hermite series and therefore a VDF cascade. Because the first term of the summation, for $m=0$, is unity, equation \ref{eqa4} can be rewritten as

\begin{equation}
    f(x) = \frac{1}{\sqrt{\pi}} e^{-\xi^2} \Biggl[ 1+ \sum_{m=1}^{\infty} \frac{H_m(\xi)}{m!} \Delta^m \Biggr].
    \label{eqa5}
\end{equation}

\noindent This expression indicates that a deviation from a Gaussian distribution can be represented as the sum of the original Gaussian core and Hermite-polynomial-like structures.

\section{Quadrature rule}\label{sec_b}

The coefficient integrals in equations \ref{eq10} and \ref{eq11} are evaluated using Gaussian quadrature. For an arbitrary function $f(x)$, its integral, $I[f]$, is approximated by a weighted sum of the function values at $N$ quadrature nodes, $Q_N[f_i]$:

\begin{equation}
    I[f] \equiv \int f(x) p(x) dx \approx \sum_{i=1}^{N} w(x_i) f(x_i) \equiv Q_N[f_i],
    \label{eqb1}
\end{equation}

\noindent where $p(x)$ is a weight function, \textit{e.g.,} $p(x)=e^{-x^2}$ for Hermite polynomials on $(-\infty,\infty)$ and $p(x)=e^{-x}$ for Laguerre polynomials on $[0,\infty)$. The quantity $x_i$ is the $i$th node, and $w(x_i)$ is the corresponding node weight. The central issue is how accurately $Q_N[f_i]$ represents $I[f]$.

If $N$ nodes and $N$ weights are chosen appropriately, the resulting $2N$ parameters allow exact integration of polynomials up to order $2N-1$. For a set of monomial basis functions $\{x^0,x^1,x^2, ..., x^m\}$, $I[f]$ can be expressed schematically as

\begin{equation}
    \int f(x) p(x) dx= \sum_{q=0}^{m} a^{q} \int x^{q}p(x)dx,
    \label{eqb2}
\end{equation}

\noindent where $a_{q}$ are arbitrary coefficients, and $m \leq 2N-1$ for $N$ nodes. Because both $I$ and $Q_N$ are linear operators as equation \ref{eqb3}, the coefficients $a_q$ can be factored out by linearity when establishing exactness for each monomial $x^q$:

\begin{equation}
    I[f] = \sum_{q=0}^{m}a_q I[x^q], \qquad Q_N[f_i] = \sum_{q=0}^{m}a_q Q_N[x_i^q].
    \label{eqb3}
\end{equation}

Any polynomial function $f(x)$ of order at most $2N-1$ can be decomposed as

\begin{equation}
    f(x) = q(x) P_N(x) + r(x),
    \label{eqb4}
\end{equation}

\noindent where $P_N(x)$ is an $N$th-order polynomial, and the polynomials $q(x)$ and $r(x)$ have orders less than or equal to $N-1$. The integral of equation \ref{eqb4} multiplied with $p(x)$ is then

\begin{equation}
    \begin{aligned}
    \int f(x) p(x) dx &= \int q(x) P_N(x) p(x) dx + \int r(x) p(x) dx \\
    &= \int r(x) p(x) dx,
    \end{aligned}
    \label{eqb5}
\end{equation}

\noindent where the first term on the right-hand side vanishes because of orthogonality. Thus, $I[f]=I[r]$.

Using equation \ref{eqb4}, $Q_N[f_i]$ can be written as

\begin{equation}
    \sum_{i=1}^{N} w(x_i) f(x_i) = \sum_{i=1}^{N} w(x_i) [q(x_i) P_N(x_i) + r(x_i)].
    \label{eqb6}
\end{equation}

\noindent To ensure that $Q_N[f_i]$ matches $I[r]$, the nodes $x_i$ must be chosen as the roots of the $N$th-order polynomial $P_N$, \textit{i.e.,} $x_i \in \{x|P_N(x)=0 \}$. For instance, $H_m(x)=0$ provides $m$ roots. The term $q(x_i) P_N(x_i)$ then vanishes, so that $Q_N[f_i] = Q_N[r_i] = I[r] = I[f]$. Thus, VDF decomposition is based on the orthogonality of the basis functions and on the selection of nodes as the roots of the quadrature polynomial. The polynomial grid is discussed in the next section.

\section{Quadrature nodes and determination methods}\label{sec_c}

The preceding discussion justifies choosing the quadrature nodes as the roots of an $N$th-order polynomial, which provides $N$ roots. The remaining question is how to determine these roots. Analytical solutions for high-order polynomials, \textit{e.g.,} $N=50$, are generally unavailable. Instead, the roots can be determined numerically. The Newton--Raphson (NR) method searches for roots iteratively from initial guesses using local gradients. Given a function $f(x)$ and an arbitrary point $x_n$, the function near $x_n$ can be approximated as

\begin{equation}
    f(x) \approx f(x_n) + f'(x_n) (x - x_n).
    \label{eqc1}
\end{equation}

\begin{figure*}[htb!]
\renewcommand{\thefigure}{C1}
\centering
\includegraphics[width=0.8\textwidth,height=0.92\textheight,keepaspectratio]{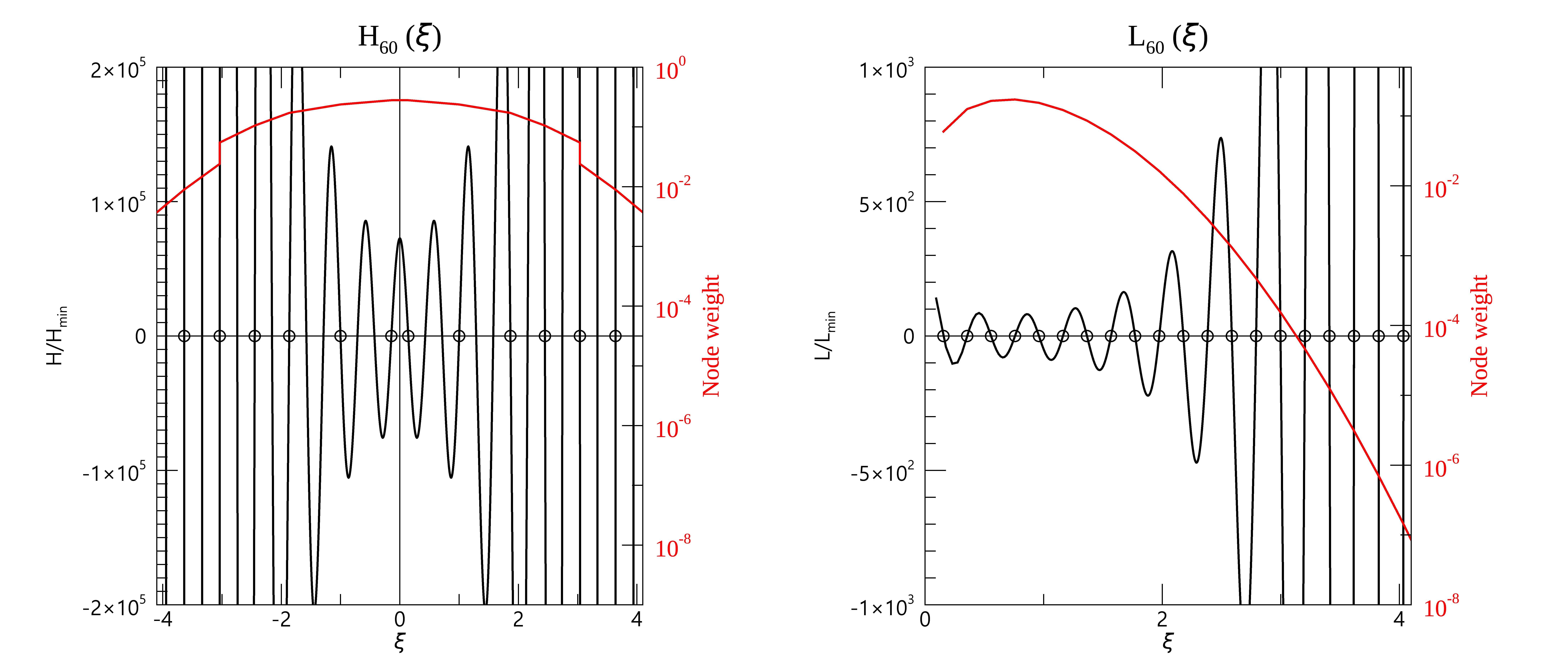}
\caption{\textit{Left panel:} Hermite polynomial of order 60, $H_{60}(\xi)$, normalized by the absolute value of its minimum, $H_{\mathrm{min}} = |\!\underset{\xi}\min H_{60}(\xi)|$, for visualization. Its roots are estimated using the Newton--Raphson method and are denoted by circles. The corresponding node weights, $w(\xi_i)$, are shown by the red solid line. \textit{Right panel:} The normalized Laguerre polynomial $L_{60}(\mu) / L_{\mathrm{min}}$ (black solid line) is displayed in $\xi$-space, where $\mu=\xi^2$, shown in the same format with its nodes (black circles) and weights (red solid line). The Laguerre nodes and weights are calculated in $\mu$ space and transformed to $\xi$ only for visualization and comparison with the linear velocity coordinates used elsewhere in this study.}\label{fig:c1}
\end{figure*}

\noindent Because $f(x)=0$ at a root,

\begin{equation}
    0 = f(x_n) + f'(x_n) (x - x_n),
    \label{eqc2}
\end{equation}

\noindent which can be rearranged as

\begin{equation}
    x_{n+1} = x_n - \frac{f(x_n)}{f'(x_n)}.
    \label{eqc3}
\end{equation}

\ The NR method searches for each root through multiple iterations until convergence and requires an initial guess. This can lead to incomplete root identification and substantial computational cost. Figure \ref{fig:c1} shows $H_{60}(\xi)$ (\textit{left panel}) and $L_{60}(\xi)$ (\textit{right panel}), together with their roots, denoted by circles, estimated using the NR method. Unlike the case of $L_{60}(\xi)$, not all roots of $H_{60}(x)$ are identified. This difference can be understood by comparing the NR update scale, $|{f(x_n)}/{f'(x_n)}|$, with the local quadrature-node spacing ($\Delta \xi ^{H}_{m}$ and $\Delta \xi ^{L}_{n}$; see Appendix \ref{sec_i}). For $H_{60}(\xi)$, the NR update scale can become comparable to or larger than the node spacing for some initial guesses. The iteration therefore may jump across neighboring roots and converge repeatedly to the same node. On the other hand, the spacing for $L_{60}(\xi)$ is generally larger than the NR scale, allowing the NR iterations to converge to all nodes.

\begin{figure*}[htb!]
\renewcommand{\thefigure}{C2}
\centering
\includegraphics[width=0.8\textwidth,height=0.92\textheight,keepaspectratio]{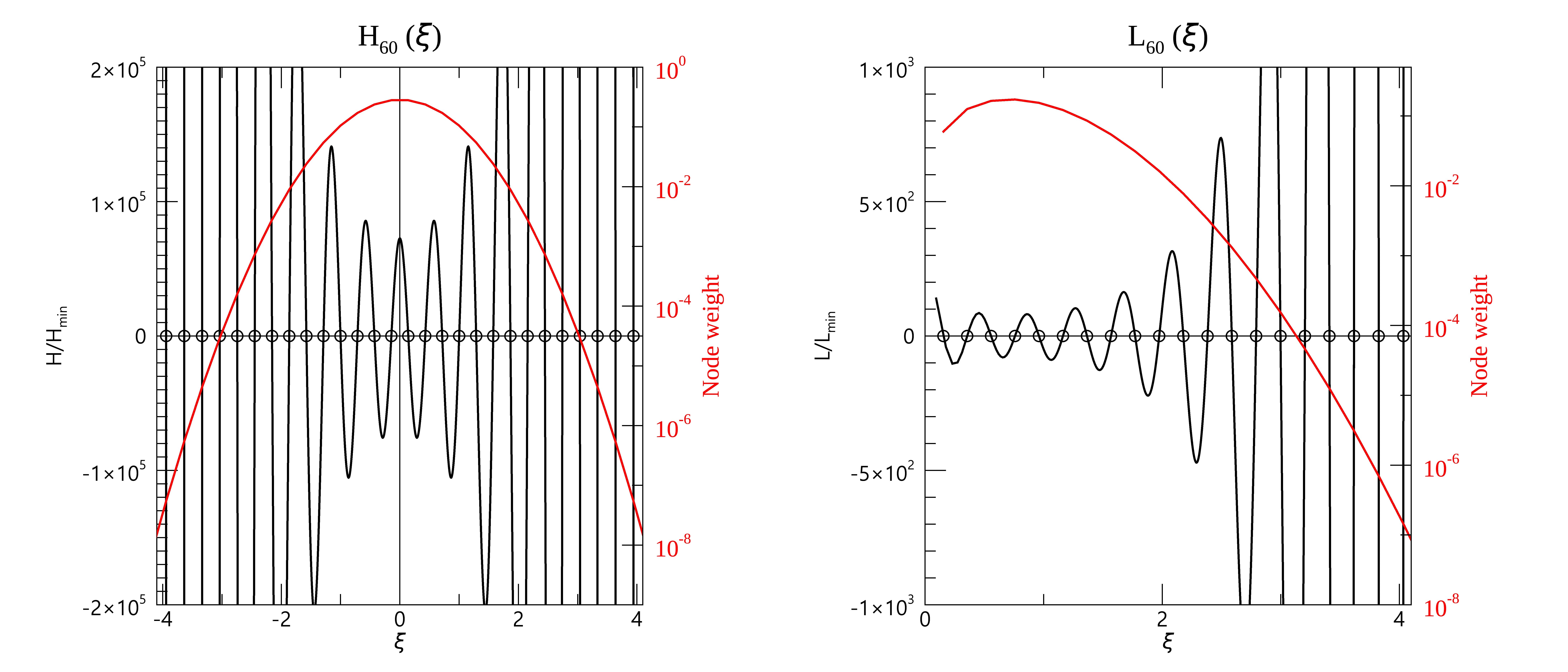}
\caption{$H_{60}(\xi)$ and $L_{60}(\xi)$, normalized by the absolute value of their respective minima, are shown in the \textit{left} and \textit{right} panels, respectively, in the same format as Figure \ref{fig:c1}. Their roots are identified using the Golub--Welsch algorithm.}\label{fig:c2}
\end{figure*}

Instead of searching for roots individually, the Golub--Welsch (GW) algorithm determines them from the eigenvalues of a matrix constructed from the polynomial recurrence relation. This algorithm identifies all roots simultaneously and is therefore more numerically stable than the NR method. For an $m$th-order polynomial $\phi_m(x)$, the three-term recurrence relation is

\begin{equation}
    x\phi_{m}(x) = a_{m+1}\phi_{m+1}(x) + b_{m}\phi_{m}(x) + a_{m}\phi_{m-1}(x),
    \label{eqc4}
\end{equation}

\noindent where $a_{m}$ and $b_{m}$ are polynomial coefficients. The Jacobi matrix $J_n$ can then be expressed as

\begin{equation}
    J_n = \begin{bmatrix}
    b_0 & a_1 & 0 & \cdots & 0 \\
    a_1 & b_1 & a_2 & \cdots & 0 \\
    0 & a_2 & b_2 & \cdots & 0 \\
    \vdots & \vdots & \vdots & \ddots & a_{n-1} \\
    0 & 0 & \cdots & a_{n-1} & b_{n-1}
    \end{bmatrix},
    \label{eqc5}
\end{equation}

\noindent so that the root-finding problem becomes an eigenvalue problem satisfying $\mathrm{det}(xI - J_n)=0$. For example, Hermite quadrature yields the Jacobi matrix

\begin{equation}
    J_n = \begin{bmatrix}
    0 & \sqrt{\frac{1}{2}} & 0 & \cdots & 0 \\
    \sqrt{\frac{1}{2}} & 0 & \sqrt{\frac{2}{2}} & \cdots & 0 \\
    0 & \sqrt{\frac{2}{2}} & 0 & \cdots & 0 \\
    \vdots & \vdots & \vdots & \ddots & \sqrt{\frac{n-1}{2}} \\
    0 & 0 & \cdots & \sqrt{\frac{n-1}{2}} & 0
    \end{bmatrix},
    \label{eqc6}
\end{equation}

\noindent based on the recurrence relation

\begin{equation}
    x\phi_{m}(x) = \sqrt{\frac{m+1}{2}}\phi_{m+1}(x) + \sqrt{\frac{m}{2}}\phi_{m-1}(x).
    \label{eqc7}
\end{equation}

\noindent Similarly, the Jacobi matrix $J_n$ for Laguerre quadrature with $k=0$ is

\begin{equation}
    J_n = \begin{bmatrix}
    1 & 1 & 0 & \cdots & 0 \\
    1 & 3 & 2 & \cdots & 0 \\
    0 & 2 & 5 & \cdots & 0 \\
    \vdots & \vdots & \vdots & \ddots & n-1 \\
    0 & 0 & \cdots & n-1 & 2n-1
    \end{bmatrix}.
    \label{eqc8}
\end{equation}

\ Figure \ref{fig:c2} shows the same polynomials as Figure \ref{fig:c1}, $H_{60}(\xi)$ and $L_{60}(\xi)$, but with their roots identified using the GW algorithm. Compared with the NR method, the GW algorithm detects all roots of both $H_{60}(\xi)$ and $L_{60}(\xi)$. We therefore use the GW algorithm for estimating the quadrature nodes. The red solid lines in both figures indicate the node weights, $w(\xi_i)$ in equation \ref{eqb1}, which are discussed in the next section.

\section{Node weight}\label{sec_d}

The right-hand side of equation \ref{eqb1}, $Q[f_i]$, consists of the values of the distribution function at the nodes $x_i$ and the corresponding weights. After determining the $N$ nodes as described in the previous section, we now estimate the node weights $w(x_i)$.

The quantity $Q[f_i]$ is a summation over discrete values. If only the discrete values of $f(x)$ at the nodes are known, \textit{i.e.,} $f(x_1)$, $f(x_2)$, $\cdots$, $f(x_N)$, the question is how a continuous approximation to $f(x)$ can be reconstructed from these values. Assuming that $f(x)$ can be represented as a polynomial, the simplest candidate reconstructed function, $t(x)$, is

\begin{equation}
    t_{N-1}(x) = \sum_{i=1}^{N} f(x_i) l_i(x),
    \label{eqd1}
\end{equation}

\noindent where the Lagrange polynomial $l_i(x)$ is

\begin{equation}
    l_{i}(x) = \prod_{\substack{j=1\\ j\ne i}}^{N} \frac{x - x_i}{x_i - x_j},
    \label{eqd2}
\end{equation}

\noindent which selects the nodes because $l_i(x_j)=\delta_{ij}$, \textit{i.e.,} $t_{N-1}(x_i)=f(x_i)$. Although the Lagrange polynomial is not always optimal for interpolation, it is adopted here for conceptual clarity.

The quadrature rule in equation \ref{eqb1} can then be written with equation \ref{eqd1} as

\begin{equation}
    \int t_{N-1}(x) p(x) dx = \sum_{i=1}^{N} f(x_i) \int l_i(x) p(x) dx.
    \label{eqd3}
\end{equation}

\noindent Comparison with $Q[f_i]$ gives

\begin{equation}
    w(x_i) = \int l_i(x) p(x) dx.
    \label{eqd4}
\end{equation}

\noindent To examine $l_i(x)$ further, consider an $N$th-order polynomial $P_N(x)$ written as

\begin{equation}
    P_N(x) = \gamma_N \prod_{j=1}^N (x - x_j),
    \label{eqd5}
\end{equation}

\noindent where $\gamma_N$ is the leading coefficient. This expression can also be written as

\begin{equation}
    \frac{P_N(x)}{x - x_i} = \gamma_N \prod_{\substack{j=1\\ j\ne i}}^N (x - x_j).
    \label{eqd6}
\end{equation}

\noindent The derivative of $P_N(x)$, denoted $P'_{N}(x)$, evaluated at $x_i$ is

\begin{equation}
    \frac{d}{dx} P_N(x) \biggr\rvert_{x=x_i} = \gamma_N \prod_{\substack{j=1\\ j\ne i}}^N (x_i - x_j).
    \label{eqd7}
\end{equation}

\noindent Combining equations \ref{eqd6} and \ref{eqd7}, the Lagrange polynomial $l_i(x)$ becomes

\begin{equation}
    l_i(x) = \frac{P_N(x)}{(x - x_i)P'_N(x_i)},
    \label{eqd8}
\end{equation}

\noindent and therefore the node weight $w(x_i)$ in equation \ref{eqd4} can be expressed as

\begin{equation}
    w(x_i) = \frac{1}{P'_N(x_i)} \int p(x) \frac{P_N(x)}{(x - x_i)} dx.
    \label{eqd9}
\end{equation}

\noindent For Hermite quadrature, where $p(x)=e^{-x^2}$ and $P_N(x)=H_N(x)$, the node weight is

\begin{equation}
    w(x_i) = \frac{1}{H'_N(x_i)} \int e^{-x^2} \frac{H_N(x)}{(x - x_i)} dx.
    \label{eqd10}
\end{equation}

\noindent Assuming that the polynomial $H_N(x)/(x - x_i)$, of order $N-1$, can be expanded as

\begin{equation}
    \frac{H_N(x)}{x - x_i} = \sum_{m=0}^{N-1} a_m H_m(x),
    \label{eqd11}
\end{equation}

\noindent then all terms except the $m=0$ term vanish by orthogonality because $H_0=1$. Specifically, the integral in equation \ref{eqd10} reduces to

\begin{equation}
    \int e^{-x^2} \frac{H_N(x)}{x - x_i} dx = \int e^{-x^2} \frac{H_N(x) H_0(x)}{x - x_i} dx =a_0 \sqrt{\pi},
    \label{eqd12}
\end{equation}

\noindent and the node weight in equation \ref{eqd10} is therefore

\begin{equation}
    w(x_i) = \frac{a_0 \sqrt{\pi}}{H'_N(x_i)}.
    \label{eqd13}
\end{equation}

\ To determine the coefficient $a_0$, we use the Christoffel--Darboux formula for a polynomial $P_m$ of order $m$,

\begin{equation}
    \sum_{m=0}^{N-1} \frac{P_m(x) P_m(y)}{n_m} = \frac{\gamma_{N-1}}{n_{N-1}\gamma_{N}} \frac{P_{N}(x) P_{N-1}(y) - P_{N-1}(x) P_{N}(y)}{x-y},
    \label{eqd14}
\end{equation}

\noindent where $n_m=\|P_m\|^2$, and $\gamma_N$ is the leading coefficient of $P_{N}$. For Hermite polynomials, $P_m$ is replaced by $H_m$, $n_m=2^m m! \sqrt{\pi}$, and the ratio of leading coefficients is $\gamma_{m-1}/\gamma_{m}=2$. Given $y=x_i$, the Christoffel--Darboux formula for Hermite polynomials becomes

\begin{equation}
    \sum_{m=0}^{N-1} \frac{H_m(x) H_m(x_i)}{2^m m!} = \frac{H_{N}(x) H_{N-1}(x_i) - H_{N-1}(x) H_{N}(x_i)}{2^N (N-1)! (x - x_i)}.
    \label{eqd15}
\end{equation}

\noindent Since $H_N(x_i)=0$, this becomes

\begin{equation}
    \frac{H_{N}(x)}{x - x_i} = \frac{2^N (N-1)!}{ H_{N-1}(x_i)} \sum_{m=0}^{N-1} \frac{H_m(x) H_m(x_i)}{2^m m!}.
    \label{eqd16}
\end{equation}

\noindent Comparison with equation \ref{eqd11} gives $a_0$ from the $m=0$ term as

\begin{equation}
    a_0 = \frac{2^N (N-1)!}{H_{N-1}(x_i)}.
    \label{eqd17}
\end{equation}

\noindent Therefore, the Hermite quadrature node weight in equation \ref{eqd13} is

\begin{equation}
    w(x_i) = \frac{2^N (N-1)! \sqrt{\pi}}{H'_N(x_i) H_{N-1}(x_i)}.
    \label{eqd18}
\end{equation}

\ The derivative of the Hermite polynomial at $x_i$ satisfies

\begin{equation}
    H_N'(x_i) = 2N H_{N-1}(x_i),
    \label{eqd19}
\end{equation}

\noindent and hence

\begin{equation}
    w(x_i) = \frac{2^{N-1} N! \sqrt{\pi}}{N^2 [H_{N-1}(x_i)]^2}.
    \label{eqd20}
\end{equation}

\noindent Similarly, the node weight for Laguerre quadrature with $k=0$ is

\begin{equation}
    w(x_i) = \frac{x_i}{(N+1)^2 [L_{N+1}(x_i)]^2}.
    \label{eqd21}
\end{equation}

\section{Interpolation of VDFs onto Quadrature Grids}\label{sec_e}

The instrument coordinate system, or the frame rotated into the MFA direction, does not necessarily coincide with the polynomial grid, \textit{e.g.,} an HH grid. Calculation of the polynomial coefficients in equations \ref{eq10} and \ref{eq11} through the quadrature rule in equation \ref{eqb1} requires the value of the distribution function at each node, \textit{i.e.,} $f(x_i,y_j)$, where $x_i$ and $y_j$ are nodes in the parallel and perpendicular coordinates of the grid, respectively. Interpolation of $f(\xi_{\parallel},\xi_{\perp})$ onto a polynomial grid is therefore required. Among several possible interpolation methods, we adopt inverse-distance weighting \citep{shepard1968}, in which the distance $r_k$ from a target point $(x_0,y_0)$ is defined as

\begin{equation}
    r_k = \sqrt{(x_k - x_0)^2 + (y_k - y_0)^2},
    \label{eqe1}
\end{equation}

\noindent where the index $k$ denotes a measured data point near $(x_0,y_0)$. The assigned weight is

\begin{equation}
    w_k = \frac{1}{r_k^p + \epsilon},
    \label{eqe2}
\end{equation}

\noindent where $p$ controls the distance weighting in the interpolation, with $p=2$ used here, and $\epsilon$ is a small positive parameter that prevents division by zero; $\epsilon < 10^{-16}$ is sufficient. The interpolated distribution function is then

\begin{equation}
    f(x_i,y_j) = \frac{\sum_{k=1}^{K} w_k f_k}{\sum_{k=1}^{K} w_k}.
    \label{eqe3}
\end{equation}

\begin{figure*}[htb!]
\renewcommand{\thefigure}{E1}
\centering
\includegraphics[width=0.99\textwidth]{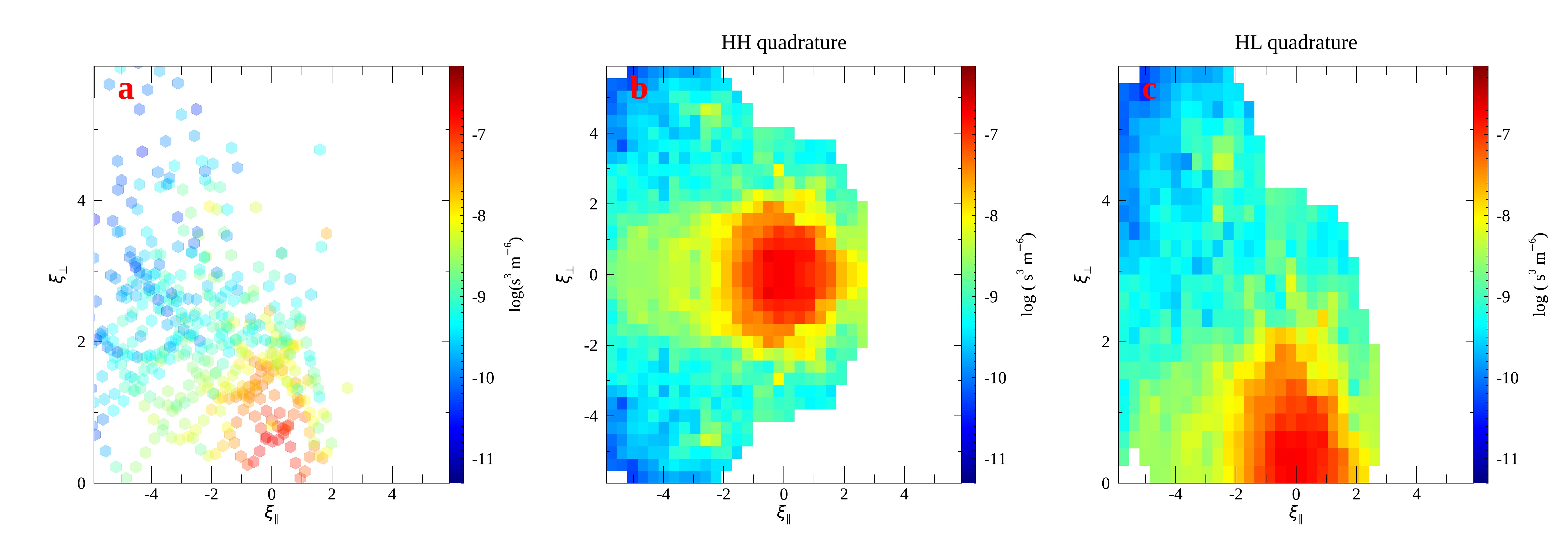}
\caption{Ion VDF observed by SWA-PAS aboard SolO at 2022 March 8 14:45:22 (a), and its interpolation onto the $60 \times 60$ (\textit{i.e.,} $M=N=60$) HH (b) and HL quadrature grids (c).}\label{fig:e1}
\end{figure*}

The number of data points used for this interpolation, $K$, should be limited to reduce computational cost, \textit{e.g.,} $K<10$. The distance $r_k$ should also be restricted, \textit{e.g.,} $r_k<0.5$, to avoid using unnecessarily distant data points. This constraint may, however, affect grid points in sparsely sampled regions. The interpolated phase-space densities (PSDs) on the HH and HL quadrature grids are shown in Figure \ref{fig:e1}, together with the PSD observed by the SWA-PAS instrument aboard SolO at 2022 March 8 14:45:22. The PSD interpolated onto the HL quadrature grid, where $M=N=60$, \textit{i.e.,} $60 \times 60$, closely resembles the observed PSD. In contrast, the HH grid of $60 \times 60$ mirrors the positive perpendicular domain into the negative domain, and the interpolated PSD is therefore symmetric about zero.

\begin{figure*}[htb!]
\renewcommand{\thefigure}{E2}
\centering
\includegraphics[width=0.99\textwidth]{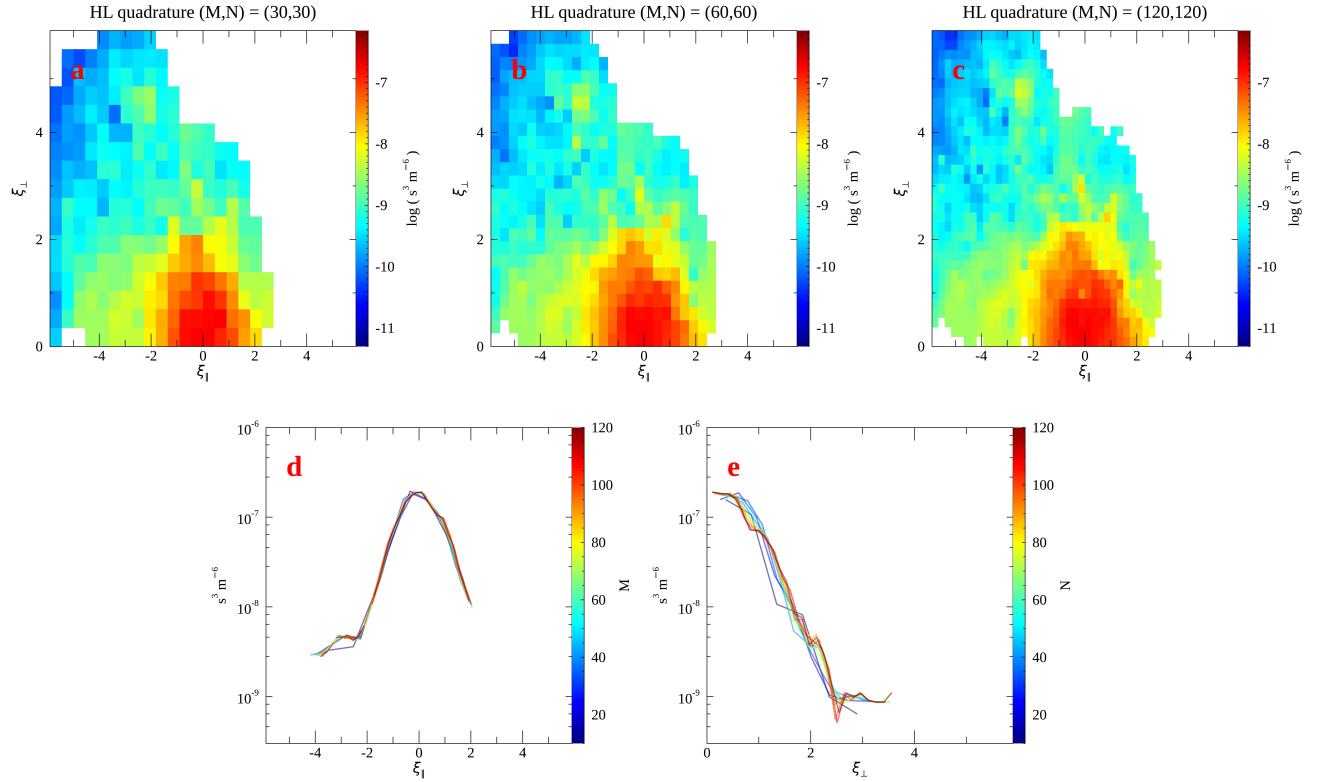}
\caption{Variation of the interpolated VDF with quadrature order. Panels show (a) $30\times30$, (b) $60\times60$, and (c) $120\times120$. Panels (d) and (e) show the parallel ($\xi_{\perp}=0$) and perpendicular ($\xi_{\parallel}=0$) cuts, respectively, of the interpolated VDF for orders from 10 to 120.}\label{fig:e2}
\end{figure*}

The choice of the number of nodes, $M, N$, is not unique. For example, \citet{larosa2025} adopt a $50\times50$ grid, yielding $2500$ quadrature nodes, comparable to the $2048$ instrument-coordinate points of SPAN-I aboard Parker Solar Probe. \citet{coburn2024} adopt a $60\times60$ grid, determined by increasing $M, N$ until the low-order spectrum, \textit{e.g.,} as shown in Figure \ref{fig:1}, becomes stable. Figure \ref{fig:e2} shows the VDF interpolated onto HL grids with different numbers of quadrature nodes, $M=N=30$, 60, and 120. The $30\times30$ grid gives a noticeably coarser representation, whereas the $60\times60$ and $120\times120$ grids yield similar VDF morphology over the well-sampled domain. The bottom panels of Figure \ref{fig:e2} show 1D cuts of the interpolated VDF in the parallel ($\xi_{\perp}=0$) and perpendicular directions ($\xi_{\parallel}=0$) with the increasing quadrature order. While the parallel VDF does not fluctuate noticeably with increasing order, the perpendicular VDF in the low-energy domain (\textit{e.g.,} $\xi_{\perp} < 2$) varies until $N$ reaches approximately 60. Therefore, $M=N=60$ is adopted throughout this study as a practical compromise between numerical convergence and computational cost. Higher quadrature orders produce a denser numerical grid, while they do not create additional independent VDF information; as discussed in Appendix \ref{sec_i}, the physically interpretable order is regulated by the instrumental velocity-space resolution.

\begin{figure*}[htb!]
\renewcommand{\thefigure}{F1}
\centering
\includegraphics[width=0.70\textwidth]{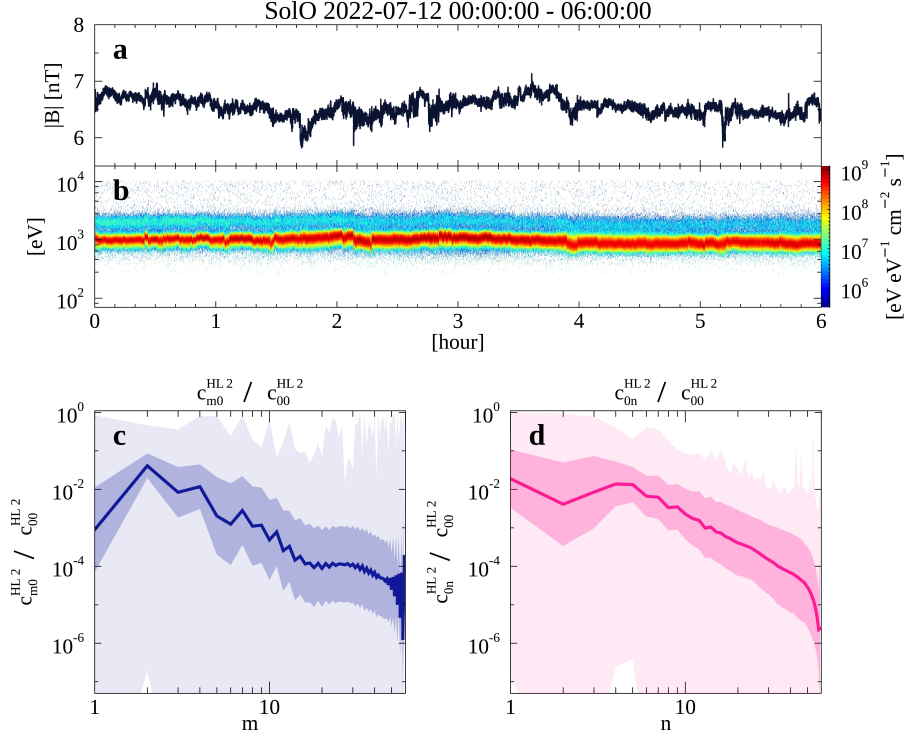}
\caption{Hermite and Laguerre spectra estimated from SolO observations at approximately 1 au during a 6 hr interval from 2022 July 12 00:00:00 to 06:00:00. Shown are (a) magnetic field, (b) ion differential energy flux, (c) normalized parallel Hermite spectra, ${c_{m0}^{HL}}^2/{c_{00}^{HL}}^2$, and (d) normalized perpendicular Laguerre spectra, ${c_{0n}^{HL}}^2/{c_{00}^{HL}}^2$, of VDFs observed during this interval. Thick solid lines in panels (c) and (d) indicate the average spectra, and darker and lighter shaded regions denote the $1\sigma$ and minimum--maximum ranges, respectively.}\label{fig:f1}
\end{figure*}

\begin{figure*}[htb!]
\renewcommand{\thefigure}{F2}
\centering
\includegraphics[width=1.0\textwidth]{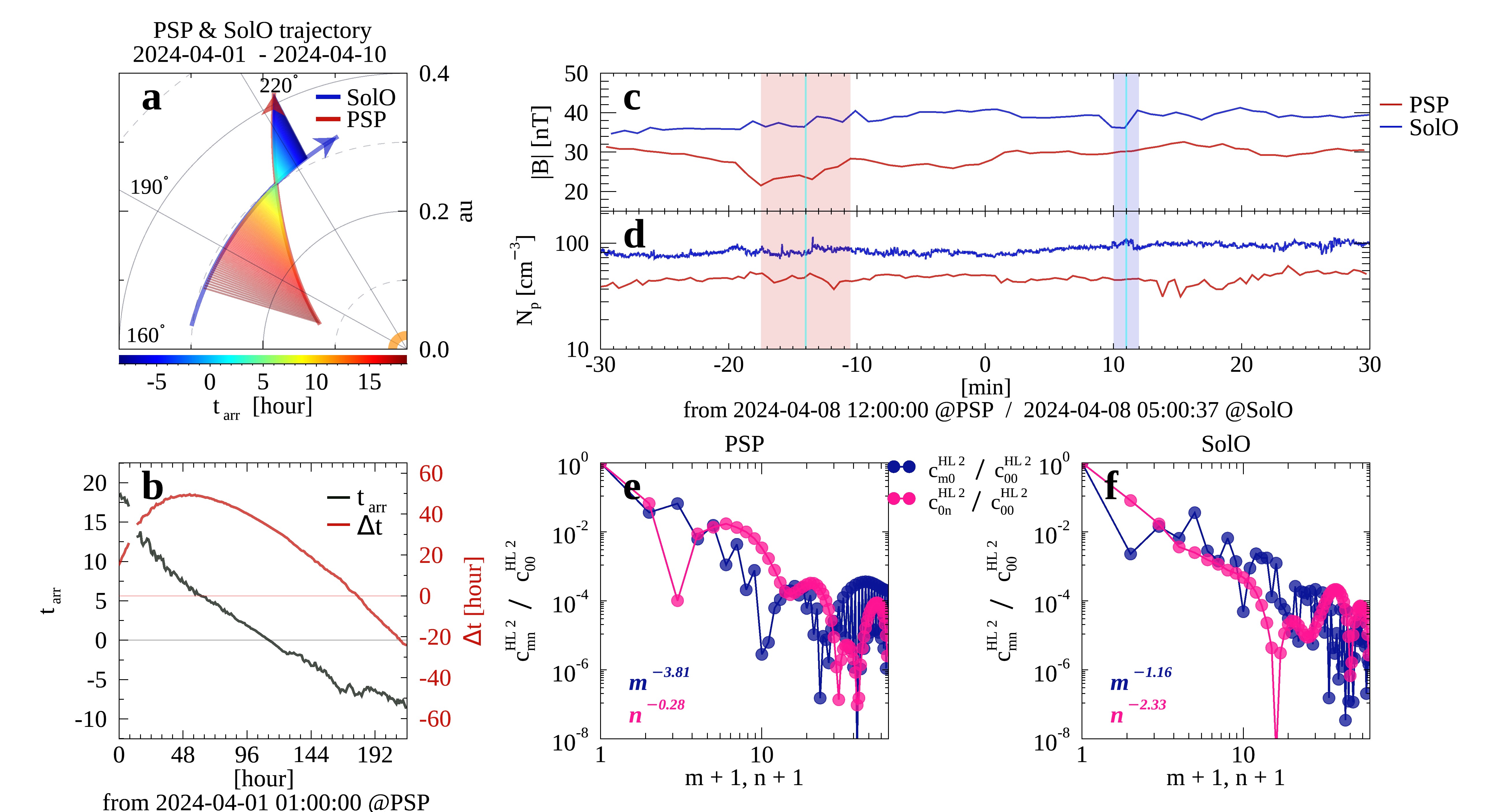}
\caption{Spectra of nearly identical propagating ion VDFs observed by PSP and SolO. (a) Trajectories of PSP and SolO in the ecliptic plane of the HGI coordinate system. The differently colored lines indicate $t_{\rm arr}$ of solar wind measured by PSP to SolO. (b) $t_{\rm arr}$ (black solid line) and $\Delta t$ (red solid line) as functions of elapsed time from 2024 April 1 01:00:00 at PSP. (c) Magnetic-field magnitude and (d) proton number density measured by PSP (red solid line) and SolO (blue solid line), respectively. The x-axis represents elapsed time in minutes from 2024 April 8 12:00:00 at PSP and 2024 April 8 05:00:37 at SolO, based on the timing analysis described in the text. The blue-shaded plasma parameters observed by SolO are expected to correspond to those in the red-shaded regions observed by PSP. The vertical cyan lines indicate the times at which the VDFs used for HL spectra were observed by PSP (Figure \ref{fig:f2}e) and SolO (Figure \ref{fig:f2}f), respectively. The blue and pink solid lines indicate ${c_{m0}^{HL}}^2/{c_{00}^{HL}}^2$ and ${c_{0n}^{HL}}^2/{c_{00}^{HL}}^2$, respectively.}\label{fig:f2}
\end{figure*}

\section{Investigation of a Potential Noise Floor in the VDF Spectrum}\label{sec_f}

To examine a potential noise floor in polynomial spectra of VDFs, we present overlapping HH and HL spectra of VDFs from SWA-PAS aboard SolO in Figures \ref{fig:f1}c and \ref{fig:f1}d, respectively, during a 6 hr interval beginning on 2022 July 12 at 00:00:00, when the solar wind was relatively quiet at approximately 1 au (see Figures \ref{fig:f1}a and \ref{fig:f1}b). The spectra in Figures \ref{fig:f1}c and \ref{fig:f1}d do not exhibit a well-defined noise floor represented as a repeatable high-order plateau, in contrast to magnetic-field spectra, for which an instrumental noise floor is often identifiable \citep{woodham2018}.

A further investigation of this issue is presented in Figure \ref{fig:f2} using PSP and SolO observations obtained at a relatively small radial separation. Figure \ref{fig:f2}a shows their trajectories during the 10 day interval from 2024 April 1 to 2024 April 10. The arrival time of solar wind observed at PSP to the SolO orbit, $t_{\rm arr}$, is defined as

\begin{equation}
    t_{\rm arr}=\frac{R_{\rm SolO} - R_{\rm PSP}}{V_{\rm R,SW}},
    \label{eqf1}
\end{equation}

\noindent where $V_{\rm r,SW}$ is the radial component of the solar wind speed measured by PSP, and $R_{\rm SolO}$ and $R_{\rm PSP}$ denote the radial locations of SolO and PSP at the time of the PSP solar wind measurement. To identify possible correspondence between plasma observed by PSP and SolO, the timing difference is defined as

\begin{equation}
    \Delta t = t_{\rm SolO} - (t_{\rm PSP} + t_{\rm arr}),
    \label{eqf2}
\end{equation}

\noindent where $t_{\rm SolO}$ and $t_{\rm PSP}$ are the SolO and PSP time stamps, respectively, during this interval. Figure \ref{fig:f2}b shows both $t_{\rm arr}$ and $\Delta t$ as functions of elapsed time from 2024 April 1 01:00:00 at PSP. A positive (negative) $t_{\rm arr}$ indicates that PSP (SolO) observes the solar wind before SolO (PSP), and a positive (negative) $\Delta t$ indicates that SolO reaches the corresponding radial location after (before) the plasma observed at PSP. Thus, $\Delta t \approx 0$ implies that PSP and SolO may observe the same plasma parcel. Based on this timing analysis and on signatures in the magnetic field (Figure \ref{fig:f2}c) and proton density (Figure \ref{fig:f2}d), we identify intervals where a nearly identical plasma is observed by SolO and PSP, shown by blue and red shaded regions, respectively. 

To assess whether a common high-order noise floor is reproduced across different spacecraft, we compare the coefficient-power spectra measured by PSP (Figure \ref{fig:f2}e) and SolO (Figure \ref{fig:f2}f) for these intervals. Fine velocity-space structure need not remain similar between the two spacecraft because kinetic processes during propagation can modify the VDF. Neither spectrum shows a clearly defined noise floor, although flattening occurs variably beyond orders of approximately 10--20. The absence of a common plateau does not imply that instrumental noise is negligible; instead, it argues against the existence of a universal, instrument-independent VDF noise floor.

This behavior is expected because the decomposition coefficients are not direct measurements of plasma noise; they depend on how each instrument samples the VDF in energy, elevation, and azimuth, including the number of bins, angular coverage, energy resolution, and mismatch between measurement and quadrature coordinates. Furthermore, the high-order behavior is controlled by the properties of the decomposition itself. For an ideal Maxwellian distribution established on decomposition coordinate, the resulting coefficient power is expected to be decrease with ascending orders, which is not generally achievable from the actual spacecraft measurements. In addition, interpolation across sparsely sampled regions, incomplete angular coverage, sharp population boundaries, and mismatch between the chosen basis and the actual VDF morphology can all generate artificial high-order power; high-order Hermite and Laguerre functions are increasingly sensitive to these small-scale irregularities. Thus, flattening of the HH spectrum near $m \approx 10$--20 is more conservatively interpreted as the onset of an effective decomposition-residual regime rather than as a direct instrumental noise floor. The absence of an equivalent clear plateau in the HL spectrum further suggests that the high-order behavior depends on coordinate representation, \textit{i.e.,} basis choice.

We note that the PSP/SPAN-I VDF in this interval is affected by limited field-of-view coverage, which likely contributes to the lower density compared with SolO. However, as the proton core remains identifiable, the polynomial decomposition can still be applied to characterize the resolved core-dominated VDF structure. Therefore, this PSP--SolO comparison is used as a representative application to a rare interval where the two spacecraft were closely aligned and likely sampled the same plasma packet, rather than as a quantitative comparison of the full proton VDF or plasma moments.

\section{Gibbs phenomenon in reconstructed VDF}\label{sec_g}

\begin{figure*}[htb!]
\renewcommand{\thefigure}{G}
\centering
\includegraphics[width=0.7\textwidth]{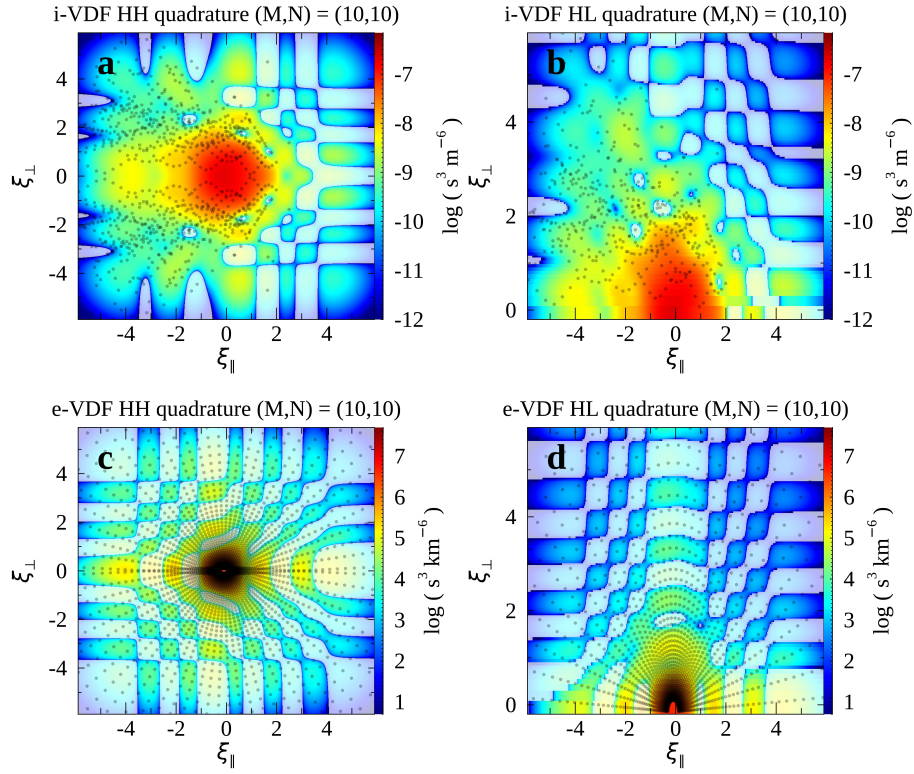}
\caption{Comparison of reconstructed VDFs with truncation order 10, \textit{i.e.,} $M=N=10$. Reconstructed ion VDFs using (a) HH and (b) HL quadrature are shown, while panels (c) and (d) show reconstructed electron VDFs using HH and HL quadrature, respectively. The black dots indicate the $\xi$-space coordinates of SWA-PAS/EAS where the ion/electron VDFs are measured, respectively. The shaded areas show the negative-valued regions from the mathematical representations.}\label{fig:g}
\end{figure*}

Reconstructed VDFs can contain negative values and ripple-like structures, as shown in Figure \ref{fig:3}. These features are analogous to the Gibbs phenomenon, in which oscillatory over- and undershoots appear near sharp gradients or discontinuities when a signal is represented by a finite Fourier series. Figure \ref{fig:g} presents ion and electron VDFs measured by SWA-PAS and SWA-EAS aboard SolO, respectively, reconstructed with HH and HL polynomials truncated at order $\leq 10$. The ion VDFs (Figures \ref{fig:g}a and \ref{fig:g}b) are reasonably recovered in regions where measured data points are present, although artificial structures appear in parts of velocity space not covered by the original measurements. These structures are therefore closely related to the incomplete velocity-space support of the PAS observations and to the global continuation imposed by the reconstruction basis.

The electron case is different: SWA-EAS provides a nearly full-sky view ($4\pi$ sr field of view; \citealt{owen2020}). Thus, one might expect the reconstructed electron VDF to be less affected by unsupported regions. However, the reconstructed electron VDFs (Figures \ref{fig:g}c and \ref{fig:g}d) still show polynomial-driven patterns across much of velocity space. Unlike the ion case, these apparent empty regions and ripple-like structures are not primarily determined by the absence of measured data points. Instead, they are caused from the finite-order representation itself. As Hermite and Laguerre functions contain positive and negative lobes, their finite sum does not guarantee positivity of the reconstructed phase-space density. Consequently, negative values, which may appear as empty regions in logarithmic plots, can be produced even within well-sampled velocity-space domains. This implies that finite-order polynomial reconstruction is a spectral projection, not an interpolation constrained to reproduce every measured or interpolated data point.

\section{3D Decomposition}\label{sec_h}

\begin{figure*}[htb!]
\renewcommand{\thefigure}{H}
\centering
\includegraphics[width=1.0\textwidth]{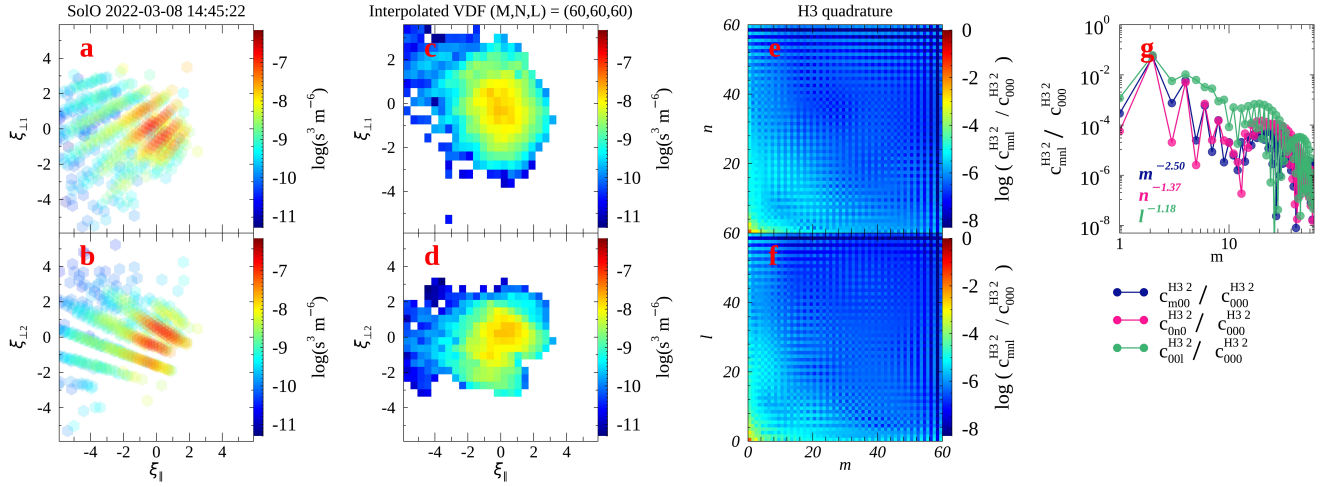}
\caption{VDF decomposition using three Hermite polynomial bases. The VDF observed by SolO at 2022 March 8 14:45:22 is shown in the velocity spaces of (a) $\xi_{\parallel}-\xi_{\perp1}$ and (b) $\xi_{\parallel}-\xi_{\perp2}$. Their interpolations onto a $60\times60$ Hermite polynomial grid are shown in panels (c) and (d), respectively. The estimated coefficients, $c_{mnl}^{H3}$, are shown in (e) $m-n$ and (f) $m-l$ spaces after averaging over the remaining component. The spectra of the respective directions, \textit{i.e.,} ${c_{m00}^{H3}}^{2}/{c_{000}^{H3}}^{2}$ (blue), ${c_{0n0}^{H3}}^{2}/{c_{000}^{H3}}^{2}$ (pink), and ${c_{00l}^{H3}}^{2}/{c_{000}^{H3}}^{2}$ (green), are plotted in panel (g), together with their low-order power laws linearly fitted within $1\leq m,n,l\leq10$, \textit{i.e.,} $m^{-2.50}$, $n^{-1.37}$, and $l^{-1.18}$.}\label{fig:h}
\end{figure*}

VDF decomposition using two orthogonal polynomial bases, such as the HH and HL representations in equations \ref{eq8} and \ref{eq9}, respectively, requires the gyrotropy assumption to project 3D distributions onto the 2D MFA frame, \textit{i.e.,} the $\xi_{\parallel}- \xi_{\perp}$ space. As the magnetic-field direction varies in the solar wind, the coordinates of the observed VDFs are generally not fixed, leading to irregular interpolation onto the 2D quadrature grid. This issue can be partially mitigated by adopting a 3D quadrature constructed from three Hermite polynomial bases, H3, as in equation \ref{eqh31}. Figure \ref{fig:h} presents an example of decomposition using H3 quadrature. Figures \ref{fig:h}a and \ref{fig:h}b show an observed VDF in the $\xi_{\parallel}- \xi_{\perp1}$ and $\xi_{\parallel}- \xi_{\perp2}$ coordinates, respectively, and Figures \ref{fig:h}c and \ref{fig:h}d show their interpolations onto the H3 quadrature grid. Compared with Figure \ref{fig:1}, the spectra in Figures \ref{fig:h}e and \ref{fig:h}f contain less pronounced high-order structure, although spectral fluctuations commonly begin near order 10 (Figure \ref{fig:h}g). This comparison confirms that high-order coefficients are sensitive to basis selection, interpolation, and mismatch between the decomposition and instrumental coordinates.

H3 is therefore useful when the available 3D measurements provide sufficient angular coverage and resolution to resolve nongyrotropic structure. However, treating the two perpendicular directions independently reduces the effective sampling density in each velocity-space dimension and may make the decomposition more sensitive to sparse coverage and measurement uncertainty. For approximately gyrotropic VDFs, the HL representation is consequently expected to remain more robust because gyrophase reduction combines measurements over the perpendicular plane and improves the effective sampling of $(\xi_{\parallel},\xi_{\perp})$ space. 

As a possible extension more naturally aligned with the angular coordinates of electrostatic particle instruments, spherical-harmonic expansions may be further considered \citep{dum1980}. Their application to spacecraft VDF decomposition will be investigated in future work.

\section{Comparison between quadrature and measurement scales}\label{sec_i}

As discussed in Section \ref{sec_e}, the selection of quadrature order in this study depends on variations of the interpolated VDFs and low-order spectra. The quadrature grid becomes denser as the order increases. However, the order does not need to be excessively large, as the quadrature-grid scale can become smaller than the physical bin size of the measuring instruments.

\begin{figure*}[htb!]
\renewcommand{\thefigure}{I}
\centering
\includegraphics[width=0.77\textwidth]{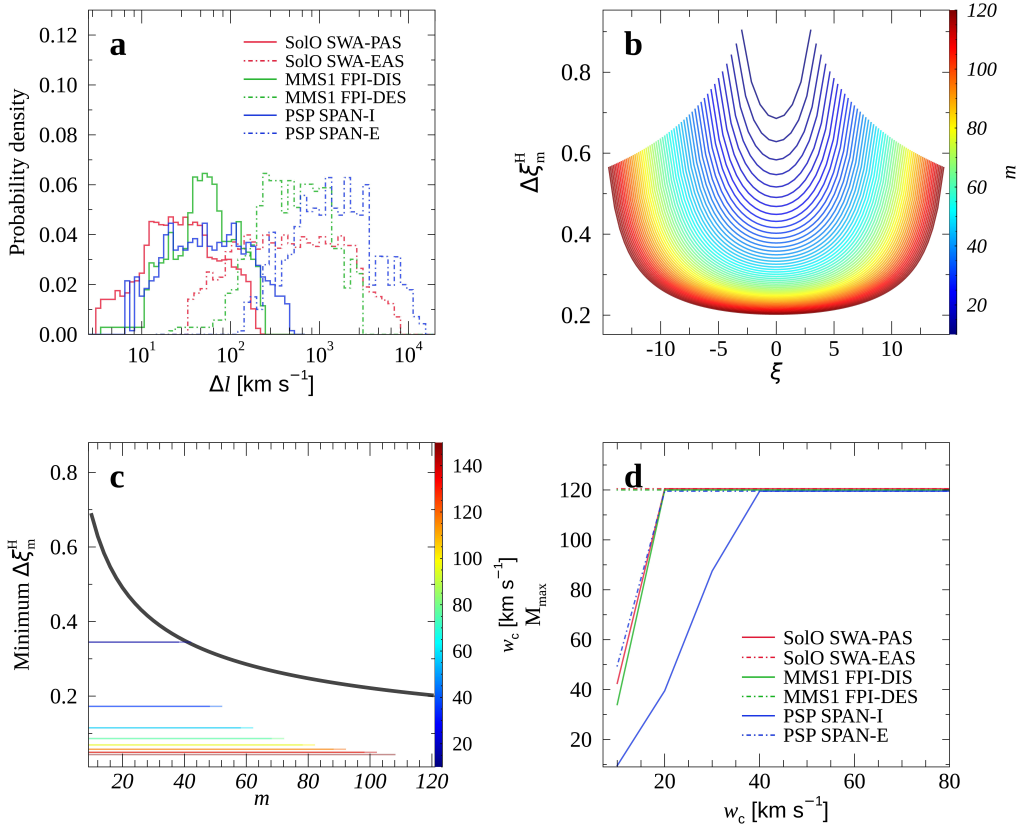}
\caption{Comparison between measurement-instrument scales and Hermite quadrature scales for decomposition. (a) Histogram of velocity-space resolution, $\Delta l$, for ion and electron instruments, including SolO SWA-PAS/EAS, MMS1 FPI-DIS/DES, and PSP SPAN-I/E. (b) Distance between adjacent Hermite quadrature nodes, $\Delta \xi_{m}^{H}$, for orders from 10 to 120 as a function of $\xi$. (c) Minimum $\Delta \xi_{m}^{H}$ (\textit{viz.}, $\!\underset{\xi}\min \Delta \xi_{m}^{H}$) as a function of order, $m$. Horizontal solid lines denote the minimum velocity-space resolution of SolO/SWA-PAS, $\!\underset{\xi}\min \Delta \xi_{\rm inst}^{PAS}$, normalized with thermal speeds from 10 to 150 km $\rm s^{-1}$. (d) Maximum quadrature order, $M_{\rm max}$, as a function of thermal speed, defined by the condition that $\!\underset{\xi}\min \Delta \xi_{m}^{H}$ remains larger than $\!\underset{\xi}\min \Delta \xi_{\rm inst}$ of the measuring instruments.}\label{fig:i}
\end{figure*}

Motivated by the search for a potential ``Nyquist frequency'' of VDF decomposition, although this quantity does not correspond to a physical frequency, we consider a wavelength in velocity space. Using the fact that the Hermite polynomial basis in equation \ref{eq3} satisfies

\begin{equation}
    \frac{\partial^2 \eta_{m}(\xi)}{\partial \xi^2} = \left( \xi^2 -2m -1  \right) \eta_{m}(\xi),
    \label{eqi1}
\end{equation}

\noindent which has the form of the time-independent Schr{\"o}dinger equation, the squared velocity-space ``wavenumber,'' $k_m^2({\xi})$, can be defined as

\begin{equation}
    k_m^2({\xi}) = 2m + 1 - \xi^2,
    \label{eqi2}
\end{equation}

\noindent where the Hermite function locally oscillates when $k_m^2({\xi})>0$ and decays exponentially when $k_m^2({\xi})<0$. The corresponding velocity-space wavelength is approximately

\begin{equation}
    l_m({\xi}) \approx \frac{2 \pi}{\sqrt{2m + 1 - \xi^2}},
    \label{eqi3}
\end{equation}

\noindent and the interval between two adjacent Hermite quadrature nodes is $\Delta \xi_{m}^{H} \approx l_m({\xi})/2$. Similarly, the Laguerre basis describes radial structure in $\mu=\xi_\perp^2$; when expressed in the linear perpendicular coordinate, its characteristic scale is approximately

\begin{equation}
    \Delta \xi_n^{L} \sim \frac{\pi}{2\sqrt{n}}.
    \label{eqi4}
\end{equation}

The Nyquist-like limit implies that $\Delta \xi_{m}^{H}$ or $\Delta \xi_{n}^{L}$ should not be smaller than the instrumental resolution, $\Delta \xi_{\rm inst}$. The Hermite scale $\Delta \xi_{m}^{H}$ reaches its minimum at $\xi=0$, which can be compared with the minimum instrumental scale, $\!\underset{\xi}\min \Delta \xi_{\rm inst}$. The highest Hermite quadrature order can then be estimated as

\begin{equation}
    M_{\rm max} \approx \frac{1}{2} \left[ \frac{\pi^2}{\left( \!\underset{\xi}\min \Delta \xi_{\rm inst} \right) ^2} -1 \right].
    \label{eqi5}
\end{equation}

\noindent For instance, when $\!\underset{\xi}\min \Delta \xi_{\rm inst} =0.2$, $M_{\rm max} \approx 123$, which is much larger than the value of 60 used in this study.

Figure \ref{fig:i} compares the scales of the Hermite quadrature and physical measuring instruments. Figure \ref{fig:i}a shows the angular and velocity-bin scales, $\Delta l$, of ion and electron instruments, including SolO SWA-PAS/EAS, PSP SPAN-I/E, and MMS1 FPI-DIS/DES. These scales are calculated as $v$, $v d\theta$, and $v \sin \theta d\phi$, where $v=\sqrt{2E/m_p}$ and $m_p$ is the proton mass for SWA-PAS, SPAN-I, and FPI-DIS and the electron mass for SWA-EAS, SPAN-E, and FPI-DES. Overall, the electron measurement bin sizes are larger than those of ion in velocity space. In addition, the measurement extents of SWA and FPI are comparable, whereas SPAN covers a higher-energy range. Figure \ref{fig:i}b presents $\Delta \xi_{m}^{H}$, showing that increasing the quadrature order from 10 to 120 produces smaller values and therefore a denser grid. Figure \ref{fig:i}c directly compares the instrumental and quadrature scales. The black solid line indicates the minimum $\Delta \xi_{m}^{H}$ (\textit{viz.}, $\!\underset{\xi}\min \Delta \xi_{m}^{H}$) of the quadrature as a function of increasing order, while the horizontal lines denote $\!\underset{\xi}\min \Delta \xi_{\rm inst}$ of SWA-PAS (\textit{viz.}, $\!\underset{\xi}\min \Delta \xi_{\rm inst}^{PAS}$) in $\xi$ space, where $\Delta \xi_{\rm inst} = \Delta l / w_c$ and $w_c$ is the proton-core thermal speed. Because $\xi$ is a velocity normalized by $w_c$ as equation \ref{vnml}, $\!\underset{\xi}\min \Delta \xi_{\rm inst}$ can be represented as a function of $w_c$, shown by different colors. For instance, $\!\underset{\xi}\min \Delta \xi_{\rm inst}^{PAS}$ --- even though its effective resolution is not a fixed quantity but varies across velocity space with the sampled energy and angular coordinates --- for the representative normalization $w_c=10 \ \mathrm{km \ s^{-1}}$ approximately corresponds to that of a Hermite quadrature order of 40, $\Delta \xi_{40}^{H}$. This implies that higher-order quadrature, \textit{e.g.,} order 60, is denser than the actual instrumental coordinates, \textit{i.e.,} $\Delta \xi_{40}^{H}>\Delta \xi_{60}^{H}$, and many interpolated values on the quadrature grid can be generated from the same encompassing data points. This can introduce high-order spectral oscillations without significantly modifying the low-order spectrum. Thus, the quadrature order need not be unnecessarily high beyond a sufficient level.

Although the limit of such a sufficient order is not fully understood, Figure \ref{fig:i}d suggests a theoretically supported quadrature order, $M_{\rm max}$, at which the quadrature scale begins to become smaller than that of the instruments. As shown in Figure \ref{fig:i}c, when $w_c>20 \ \mathrm{km \ s^{-1}}$, the quadrature-grid scale remains larger than the minimum SWA-PAS bin scale up to order 120, the maximum order considered here. Similarly, the scales of other instruments can be compared in Figure \ref{fig:i}d, where those of the electron instruments use thermal speeds multiplied by a factor of $\sqrt{m_p/m_e} \approx 42.8$ for normalization to $\xi$ space. For example, $w_c=10 \ \mathrm{km \ s^{-1}}$ for ions approximately corresponds to $428.5 \ \mathrm{km \ s^{-1}}$ for electrons. Overall, $M_{\rm max}$ reaches 120 rapidly for ion instruments when $w_c>20 \ \mathrm{km \ s^{-1}}$, whereas PSP SPAN-I requires a higher thermal speed, $w_c\approx40 \ \mathrm{km \ s^{-1}}$, to approach order 120. In contrast, the electron instrument scales are mostly smaller than the quadrature scales for thermal speeds above $428.5 \ \mathrm{km \ s^{-1}}$, whereas $M_{\rm max} \approx 50$ at this thermal speed for PSP SPAN-E and rapidly reaches 120 when the thermal speed is $857 \ \mathrm{km \ s^{-1}}$. Therefore, the use of quadrature order 60 throughout this study is reasonable for most solar wind conditions.

\end{appendix}

\bibliography{Park_bib}{}
\bibliographystyle{aasjournal}

\end{document}